\documentclass[sigconf]{acmart} 

\AtBeginDocument{%
  \providecommand\BibTeX{{%
    \normalfont B\kern-0.5em{\scshape i\kern-0.25em b}\kern-0.8em\TeX}}}

\usepackage{fancyvrb}
\copyrightyear{2026}
\acmYear{2026}
\setcopyright{cc}
\setcctype{by-nc-nd}
\acmConference[CHI '26]{Proceedings of the 2026 CHI Conference on Human Factors in Computing Systems}{April 13--17, 2026}{Barcelona, Spain}
\acmBooktitle{Proceedings of the 2026 CHI Conference on Human Factors in Computing Systems (CHI '26), April 13--17, 2026, Barcelona, Spain}
\acmDOI{10.1145/3772318.3790461}
\acmISBN{979-8-4007-2278-3/2026/04}

\usepackage{listings}
\usepackage{subfig}
\usepackage{color, soul}
\usepackage[font={scriptsize,sf},tableposition=top]{caption}
\usepackage{float}
\floatstyle{plaintop}
\restylefloat{table}
\usepackage{multirow}
\usepackage{colortbl}
\usepackage{tabularx}
\usepackage{balance}
\usepackage[utf8]{inputenc}
\usepackage[T1]{fontenc}
\usepackage{lipsum}
\usepackage{hyperref}
\usepackage[normalem]{ulem}
\usepackage{microtype}
\begin{document}

\title[Rememo]{Rememo: A Research-through-Design Inquiry Towards an AI-in-the-loop Therapist’s Tool for Dementia Reminiscence}

\author{Celeste Seah}
\orcid{0009-0002-0897-526X}
\affiliation{%
  \department{Division of Industrial Design}
  \institution{National University of Singapore}
  \city{Singapore}
  \country{Singapore}
}
\email{celestes@nus.edu.sg}

\author{Yoke Chuan Lee}
\orcid{0009-0007-5460-5688}
\affiliation{%
  \institution{ECON Healthcare Group}
  \city{Singapore}
  \state{Singapore}
  \country{Singapore}
}
\email{yokechuan@econhealthcare.com}

\author{Jung-Joo Lee}
\orcid{0000-0002-7414-1936}
\affiliation{%
  \department{Division of Industrial Design}
  \institution{National University of Singapore}
  \city{Singapore}
  \country{Singapore}
}
\email{jjlee@nus.edu.sg}

\author{Ching-Chiuan Yen}
\orcid{0000-0003-4325-1689}
\affiliation{%
  \department{Division of Industrial Design}
  \institution{National University of Singapore}
  \city{Singapore}
  \state{Singapore}
  \country{Singapore}
}
\affiliation{%
  \department{CUTE Center}
  \institution{National University of Singapore}
  \city{Singapore}
  \state{Singapore}
  \country{Singapore}
}
\email{didyc@nus.edu.sg}

\author{Clement Zheng}
\orcid{0000-0003-0336-6430}
\affiliation{%
  \department{Division of Industrial Design}
  \institution{National University of Singapore}
  \city{Singapore}
  \country{Singapore}
}
\affiliation{%
  \department{CUTE Center}
  \institution{National University of Singapore}
  \city{Singapore}
  \country{Singapore}
}
\email{didzzc@nus.edu.sg}

\renewcommand{\shortauthors}{Seah et al.}

\newcommand{\anonCountry}[0]{Singapore}
\newcommand{\anonCountrytwo}[0]{Malaysia}
\newcommand{\anonCountrythree}[0]{Philippines}
\newcommand{\anonCareOrganization}[0]{\textit{ECON Healthcare}}
\newcommand{\irbone}[0]{\textit{NUS-IRB-2024-818}}
\newcommand{\irbtwo}[0]{\textit{NUS-IRB-2025-99}}
\newcommand{\quoted}[1]{\textcolor{gray}{\textit{``#1''}}}
\definecolor{myblue}{HTML}{2853FF}
\newcommand{\prompted}[1]{\textcolor{myblue}{\texttt{``#1''}}}

\newcommand\change[1]{\textcolor{ForestGreen}{#1}}
\newcommand\remove[1]{\textcolor{red}{\sout{#1}}}

\newcolumntype{L}[1]{>{\raggedright\let\newline\\\arraybackslash\hspace{0pt}}m{#1}}
\newcolumntype{C}[1]{>{\centering\let\newline\\\arraybackslash\hspace{0pt}}m{#1}}

\renewcommand{\sectionautorefname}{Section}
\renewcommand{\subsectionautorefname}{Subsection}
\renewcommand{\subsubsectionautorefname}{Subsubsection}
\renewcommand{\figureautorefname}{Figure}
\renewcommand{\tableautorefname}{Table}


\begin{teaserfigure}
    \centering
    \includegraphics[width=\textwidth]{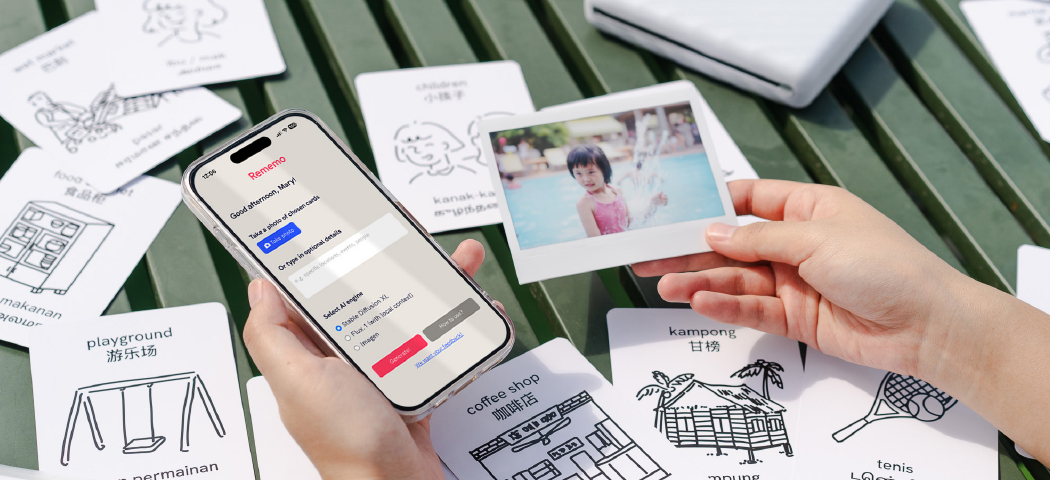}
    \Description[Rememo interface and a small printed photo]{Hands hold a smartphone showing the Rememo interface and a small printed photo of a child in a pool. On the table are black-and-white prompt cards with simple drawings and bilingual labels such as ``playground'', ``coffee shop'', and ``kampong'', illustrating how cards and the app are used together for reminiscence.}
    \caption{Rememo webapp interface on a mobile device along with illustrated prompt cards and generated image print-out.}
    \label{fig:teaser}
\end{teaserfigure}

\begin{abstract}
Reminiscence therapy (RT) is a common non-pharmacological intervention in dementia care. Recent technology-mediated interventions have largely focused on people with dementia through solutions that replace human facilitators with conversational agents. However, the relational work of facilitation is critical in the effectiveness of RT. Hence, we developed Rememo, a therapist-oriented tool that integrates Generative AI to support and enrich human facilitation in RT. Our tool aims to support the infrastructural and cultural challenges that therapists in \anonCountry\ face. In this research, we contribute the Rememo system as a therapist’s tool for personalized RT developed through sociotechnically-aware research-through-design. Through studying this system in-situ, our research extends our understanding of human-AI collaboration for care work. We discuss the implications of designing AI-enabled systems that respect the relational dynamics in care contexts, and argue for a rethinking of synthetic imagery as a therapeutic support for memory rather than a record of truth.
\end{abstract}





\maketitle
\section{Introduction}\label{sec:introduction}
Populations worldwide are aging rapidly, and the widening gap between lifespan and healthspan means that older adults are increasingly spending their later years in poor health \cite{garmany2021}. Dementia, a family of neuro-degenerative diseases, is among the most prevalent conditions \cite{who} contributing to shaping this trajectory. As dementia progresses, cognitive function declines and is often accompanied by changes in emotions, behavior, and motivation. At different stages of dementia, different interventions based on person-centered care are required to support persons living with dementia. Reminiscence therapy (RT) is one of the most widely used psychosocial interventions to support this process, particularly for individuals in the early to middle stages. By engaging people living with dementia in the discussion of past experiences, RT aims to stimulate memory recall, evoke positive mood, and foster a sense of personal identity \cite{macleod2020}.

RT often involves the use of memory triggers such as photographs, objects, music, or scents that engage multiple senses. Among these, visual triggers have been shown to be particularly effective, as episodic memories are encoded through visual imagery \cite{conway2009, greenberg2005}. Personal photographs are commonly used in informal care, but access to them is limited in institutional settings. In such contexts, therapy staff turn to curated image libraries or archives, which studies suggest can elicit rich, emotionally significant storytelling, sometimes even more so than personal photographs \cite{astell2010}. Yet the task of sourcing images that are culturally and personally resonant for people living with dementia remains labor-intensive for therapy staff preparing sessions.

Over the past two decades, HCI research has sought to address these challenges through technology-mediated interventions. Existing approaches range from recommender systems that suggest materials based on certain indicators, to conversational agents that seek to replicate the role of the facilitator by engaging users directly. While such systems demonstrate advanced technical capabilities, they often overlook the realities of care contexts. In practice, care technologies are almost always set up, mediated, and overseen by professional staff. Moreover, practitioners consistently emphasize that technological systems cannot replace the relational and affective labor of facilitation, particularly for a vulnerable group such as individuals with cognitive impairments. At the same time, the aged care sector faces persistent structural challenges: chronic under-staffing, high dependence on migrant workers, and frequent language and cultural barriers between staff and residents. In multilingual, multicultural contexts such as \anonCountry, these challenges are especially pronounced.

In this paper, we present findings from a two-year research-through-design (RtD) project conducted in collaboration with occupational therapy teams to explore how Generative AI might productively support RT facilitation. The outcome of this collaboration is Rememo, a therapist-oriented tool that leverages generative image models to personalize visual memory triggers while maintaining therapist oversight and resident agency. We deployed Rememo in a two-week field study with 5 care staff across two private nursing homes in \anonCountry, involving 21 residents and producing 151 generated images.

Our findings demonstrate how Rememo eased session preparation and helped bridge communication barriers. Participants reported that the system elicited richer recall and more sustained conversation among residents, contributing to positive mood and engagement. On the other hand, the system still required adaptation to residents’ individual profiles, which indicate the key areas of therapist-AI collaboration in AI-supported RT. The deployment also surfaced frictions, such as latency in image generation, that were overlooked during the design phase and lab tests, which underscored the importance of deep RtD collaboration in the development of novel technologies. 

We discuss two central contributions of this work. First, we advance an AI-in-the-loop model \cite{green2019aitl, natarajan2025aitl} of therapist–AI collaboration, where human facilitators remain in control of the therapeutic process while AI supports at appropriate junctures to extend their impact. Second, we argue for a reframing of synthetic imagery in RT: not as a substitute for authentic photographs, but as a support for reconstructive memory work for therapeutic benefit. Together, these contributions underscore the importance of designing AI-enabled systems that respect the relational dynamics and labor conditions of dementia care, while probing new possibilities for human-centered AI in therapeutic practice.

\subsection{Motivation}\label{sec:motivation}
Initial field visits were conducted in a sensitizing capacity to better understand the material, relational, and infrastructural contexts in which RT takes place. These included informal engagements at five dementia care facilities in \anonCountry, comprising nursing homes, rehabilitation centers, and senior daycare centers. During these visits, the first author engaged in ethnographic observations and informal conversations with therapists and therapy teams. 

The first author was drawn to the metaphor of the brain being a computer and the synaptic-like structure of neural networks, she was motivated to explore the intersection of memory and AI. The first author thus established a collaboration with the second author, who is a Principal Occupational Therapist at a nursing home, to explore how AI might be productively utilized in care practices. As a means to bridge our disciplines and collectively co-create new possibilities, we adopted a sociotechnically-aware RtD methodology to develop the designed intervention. 
\section{Background}\label{sec:background}
\subsection{Cognition \& memories}\label{sec:cognition_memories}
Dementia is a family of chronic neuro-degenerative diseases that often manifests as memory loss, speech impairment and emotional disturbances \cite{cloak2024}. It is non-curable and characterized by progressive degeneration of the brain, accounting for 28.8 million Disability Adjusted Life Years worldwide in 2016. The rate of decline varies from person to person as some live with the disease for decades while others deteriorate rapidly. In the early stages, people living with dementia experience lapses in episodic memory and impairments in their ability to encode, store and retrieve new or autobiographical information. As the disease advances, semantic memory declines, communication skills and language fluency deteriorate, causing a loss of understanding and meaning of the world as a whole. At the same time, people with dementia often retain past memories and can continue to participate meaningfully when supported through a strengths-based approach \cite{macleod2020}.

To cope with increased care needs, many people living with dementia access services at dedicated dementia care centers in the community or reside in long-term residential facilities. These facilities provide pharmacological and non-pharmacological treatments to manage symptoms. In particular, non-pharmacological treatments primarily serve to improve mood while engaging in a reflective accounting of one’s experiences as a life well-lived \cite{erikson1994}.

\subsection{Reminiscence therapy}\label{sec:reminiscence_therapy}
Reminiscence therapy (RT) is a common non-pharmacological, psychosocial intervention for cognitive impairments like dementia to practice communication skills and promote active recall to maintain cognition. It has also shown to be beneficial for emotional well-being by promoting communication and interaction \cite{woods2018} and helping to manage behavioral and psychosocial symptoms of dementia. RT is an “intuitive process whereby person looks back and reflects on their life” \cite{butler1963}, underpinned by person-centered care principles \cite{kitwood1997} in negotiating between ego integrity and despair \cite{erikson1994}.

RT involves the discussion of past experiences and events between individuals or with a facilitator, and can take on different forms depending on the individual’s needs and abilities. It is person-centric and relationship driven, often mediated by memory triggers such as old photographs, physical artifacts, music, arts and crafts, activities, scents and more. The relevance of the memory trigger to the individual is crucial in determining the extent of recall and overall effectiveness of the reminiscence \cite{brooker2016}. Therapy staff also seek to understand their clients’ cognitive abilities to tailor activities that are appropriate for their level, minimising their `exposure to a failure experience’, given that a level of cognitive ability is required to partake in RT \cite{bender1998therapeutic}. 

From our primary research with care organizations in \anonCountry, we observed that RT is a fluid and relational practice with no fixed model. The therapy teams we worked with often adapted RT activities to meet the cognitive and social engagement needs of their clients. For example, a therapy staff might organize clients with similar interests together for group therapy to inspire greater social interactions. This corroborates with a narrative analysis of RT by Macleod et al. that showed that there was ``no consistency in the model or programme used’’ among the studies examined. In this research, we focus on developing a RT tool specific to local needs by adopting a RtD process in close partnership with local stakeholders.

\subsection{Labor Realities of Care Work}\label{sec:labor_realities_of_care_work}
Reminiscence facilitation is a human practice that necessitates attention to the structural conditions under which such work takes place. In many aged care systems, including \anonCountry, chronic understaffing and lack of manpower place significant strain on care delivery. The sector is heavily reliant on migrant labor, with up to 50\% of therapy aides hailing from other countries \cite{LIEN2018}. While these workers form the backbone of day-to-day operations in residential and community care facilities, they often face language barriers and unfamiliarity with the local culture.

These structural conditions present particular challenges for delivering effective RT. As discussed, facilitation relies on subtle, culturally nuanced interactions that are difficult to script or automate. Yet, some foreign-born care workers lack fluency in the local languages and dialects that residents speak. They may also be unfamiliar with specific historical or cultural references that resonate with the local population. These gaps limit their capacity to improvise and respond meaningfully to client narratives and can constrain the depth of emotional connection required for effective reminiscence.

Our fieldwork suggests that any viable technological intervention in this space must contend with the realities of who is performing the care work, under what conditions, and with what forms of support. Rather than displacing workers outright, design efforts might be more productively directed towards empowering novice facilitators through technology tools to grow their therapy expertise.
\section{Related work}\label{sec:related_work}
We conducted a pilot landscape review of literature on technology-mediated non-pharmacological interventions for RT with older adults in the last 25 years. Our aim was to understand the prior approaches to the problem space and the corresponding technological approaches from past to present, with an interest in the recent surge of AI methods. An initial search with the following search syntax on ACM Digital Library yielded 573 records. The search string was crafted to capture relevant literature that included both reminiscence concepts and the target population of older adults with or without various impairments combined with application of a technology system. After screening, removing duplicates, and snowball sampling, left 54 papers that proposed novel technological systems for reminiscence therapies with older adults and people with dementia. The full corpus can be found in the Supplementary Material.

\begin{figure*}[ht]
    \includegraphics[width=0.9\textwidth]{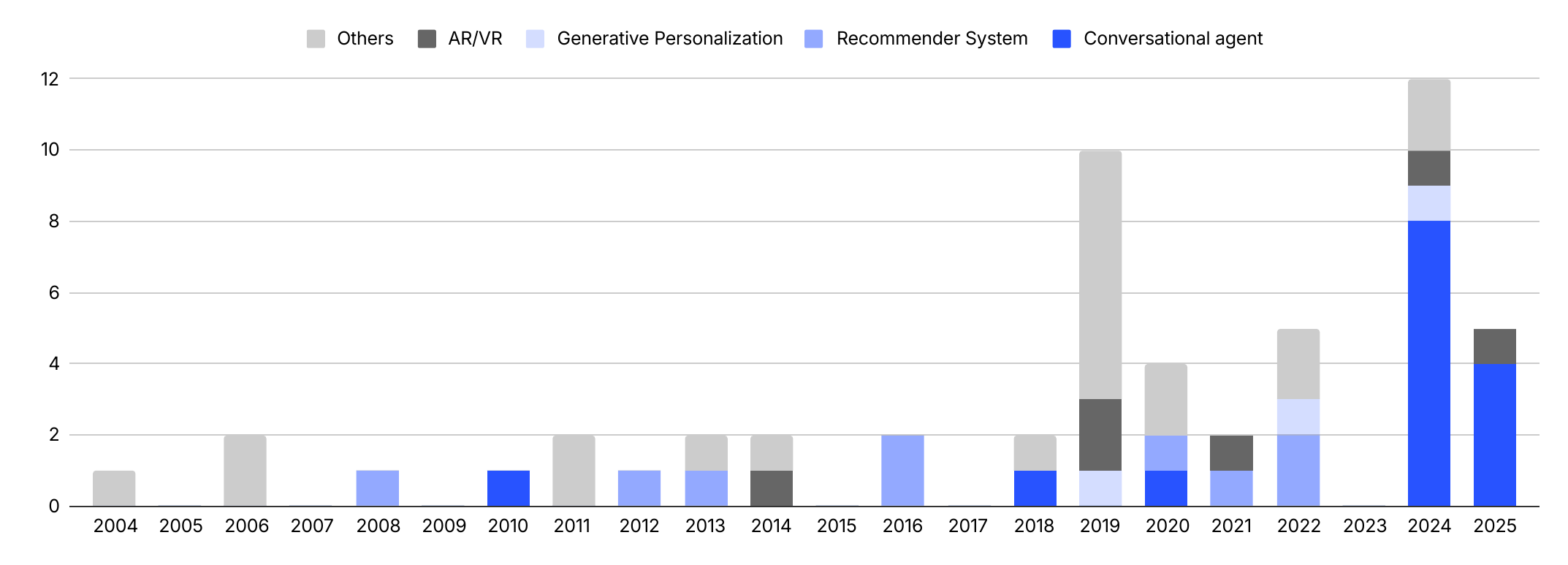}
  \caption{Increase in number of relevant studies in 2019 and 2024, with recent growth focusing on conversational agent approaches}
  \Description{Stacked bar chart showing yearly counts from 2004 to 2025 across five categories: Others, AR/VR, Generative Personalization, Recommender System, and Conversational agent. Counts are near zero through the mid-2010s, then increase. 2020 shows a spike dominated by Others. From 2022 onward, Conversational agent becomes the largest share, with the tallest bar in 2024 and small contributions from Generative Personalization and Others; 2025 remains high relative to years before 2020. Generative Personalization appears only in recent years as a small portion.}
  \label{fig:studiesgraph}
\end{figure*}

\begin{Verbatim}[xleftmargin=2em]
AllField:(reminisc* OR "life story" OR 
"life review") 
AND AllField:(therap* OR care* OR memor*) 
AND AllField:(dementia OR "older adult*" OR elder* 
OR Alzheimer* OR "cognitive impairment" 
OR "visual impairment") 
AND AllField:(AI OR "artificial intelligence" 
OR technolog* OR chatbot OR recommender 
OR immersive OR VR OR AR OR "machine learning")
&&         Research Article
\end{Verbatim}

\subsection{Technologies for RT}
There has been a marked increase in research into augmenting RT with digital technologies since 2019, with 64\% of the shortlisted studies being published in the last 7 years. This indicates growing interest in the area. These studies utilize technologies like virtual reality \cite{ppali2025care}, facial expression recognition \cite{nebot2022longremi}, knowledge graphs \cite{wang2024goodtimes}, and large language models (LLMs) \cite{sun2025reminibuddy}. 

Half of the papers examined utilized some form of AI (represented in shades of blue in \autoref{fig:studiesgraph}) in line with mainstream proliferation of NLP and LLM tools like ChatGPT and Midjourney since 2019. The broader discourse surrounding the relationship between humans and AI has evolved in tandem as well. Earlier work focused on a singular interaction level for AI implementation \cite{amershi2019guidelines} to balance autonomy between user and system \cite{horvitz1999mixed} under the umbrella of Human-AI interaction (HAII). This has widened towards embedding AI in larger service systems and workflows for Human-Centred AI (HCAI), especially in complex systems like healthcare \cite{ulloa2022invisible} where the sociotechnical infrastructure is instrumental to its success. At the same time, we have witnessed the rapid adoption of Generative AI in commercial products across sectors, raising concerns that governance policies on safety, ethics and autonomy have not kept pace \cite{wef2024governance}.

Within the papers examined, we organized the AI-enabled projects into three approaches:

\begin{enumerate}
    \item \textit{\textbf{Conversational Agents:}} Systems designed to replace or replicate the role of a human facilitator by directly engaging the user in dialogue about their past. The user interacts primarily with the system, which absorbs the roles of both facilitator and stimuli. Note that some of these systems also incorporate generative personalization capabilities as a secondary function. (e.g., Remihaven \cite{zhang2025remihaven}, GoodTimes \cite{wang2024goodtimes}, Elisabot \cite{caros2020elisabot}).
    \item \textit{\textbf{Recommender Systems:}} Systems that curate and suggest pre-existing reminiscence materials based on certain criteria, but the actual reminiscence and discussion occurs outside the system. (e.g., LONG-REMI \cite{nebot2022longremi}, MomentMeld \cite{kang2021momentmeld}, Photostroller \cite{gaver2011photostroller}).
    \item \textit{\textbf{Generative Personalization:}} Systems that use Generative AI to create novel, personalized stimuli from user input or life history, though the resulting reflection may occur either with or without system involvement. (e.g., Musical AIs for Musical Therapy \cite{sun2024music}).
\end{enumerate}

These projects show the potential benefit that AI can bring to both caregivers and care recipients. For facilitators, it can curate relevant material (Recommender Systems) or even conduct RT independently (Conversational Agents). For persons living with dementia, it can customize stimuli to each individual’s experiences (Generative Personalization).

Generative Personalization in particular is especially relevant to technology-mediated RT as it supports  user engagement and personalization \cite{zhang2025remihaven}. AI-based music therapy tools were perceived to be promising in increasing the efficiency of RT, enriching and personalizing content for improved client engagement in Sun et al.’s work \cite{sun2024music}. For image-based RT, facilitators are no longer constrained by the choice between generic photographs and personal photographs \cite{astell2010} with the advent of Generative AI tools that can produce content tailored to each individual’s experiences. The scarcity of personal photos and impersonal nature of generic cues are effectively eliminated with Generative AI as demonstrated by Nan et al. \cite{nan2025kimono}, Remihaven \cite{zhang2025remihaven}, and Zhai et al. \cite{zhai2024chatgpt}. Yet, Generative Personalization as an approach remains relatively underexplored in comparison to other AI-driven approaches of Conversational Agents and Recommender Systems (\autoref{fig:studiesgraph}).

Of the 54 studies reviewed, 47 (87\%) targeted the individual and 45 (83\%) focused on creating stimuli for RT. In contrast, only 16 (30\%) explicitly involved RT facilitators, despite their central role in the delivery of dementia care. We believe that technological interventions ought to be designed for, and \emph{with}, this group. Sun et al’s work highlights how individuals with dementia face challenges using and engaging directly with most technology-mediated RT systems due to cognitive challenges, lack of trust, and safety concerns \cite{sun2024tmnpi}. They also show how such intervention technologies often display a lack of understanding of the professional care practice and are not designed to support therapy staff in their work. This broader trend treads an uneasy tension between supporting and supplanting therapy staff \cite{sun2024music} and the overwhelming majority of the Conversational Agent approaches in the surveyed literature shows a continuation of this trajectory.

There are tangible benefits to interventions that include facilitators as a key actor such as stimulating richer storytelling \cite{siriaraya2014worlds} and mitigating risk of discomfort \cite{waycott2022role}. We explore how and why facilitation is a fundamentally human practice of connection between the individual and the facilitator.

\subsection{Therapy facilitation as a human practice}\label{sec:therapy_facilitation_as_a_human_practice}
Care technologies are frequently conceptualized in terms of their benefit to care recipients while rendering care workers as peripheral users or passive intermediaries in the system's design and evaluation ~\cite{sun2023data}. However, in practice, it is the care workers who serve as the primary interface between such technologies and older adults. Waycott et al.’s work frames care workers as “gatekeepers” by electing which individuals would or would not be suitable for particular technologies due to their prognosis \cite{waycott2022role}. The actions, interpretations, and improvisations of these workers significantly shape how these technologies are enacted in everyday care settings, ultimately determining the success or failure of therapy.

In the domain of technology-mediated RT, this disconnect is particularly salient as recent innovations tend to overlook the human facilitator. Many conversational agent tools aim to supplant the human facilitator, such as implementing multi-agents to achieve the effect of varying personas \cite{zhang2025remihaven, sun2025reminibuddy}. While such tools are useful for independent reminiscence when conversational partners are not available, they risk oversimplifying the therapeutic process, ignoring the complex social and emotional dynamics involved in memory work. Emotional sensitivity is a weakness of digital systems with no human oversight \cite{xygkou2024mindtalker}. As Hsu et al. note, such systems are often ill-equipped to navigate the affective nuances of aging, loss, and identity that arise in reminiscence contexts \cite{hsu2025bittersweet}.

Effective reminiscence facilitation is highly relational and situated, demanding not only practical knowledge of therapy goals, but also adaptive communication and empathic listening to create a supportive social environment for each individual \cite{brooker2016}. Bender et al. give concrete guidance for facilitators to adapt RT to different individuals, such as seating clients to align personalities and minimize disruptive behaviors \cite{bender1998therapeutic}. Our fieldwork echoed this adaptivity as therapy staff frequently described facilitation as a dynamic, improvisational skill. These relational competencies are rarely formalized in training programs, often acquired tacitly over time through experiential learning and affective labor. While large language models (LLMs) can be designed and prompted to ask relevant questions, they do not possess the capacity to interpret subtle cues, shift conversational tone, or express genuine affect. For novice facilitators, this makes reminiscence a similarly challenging practice to enter as they may struggle to communicate and connect with their clients. Foregrounding the role of facilitation as a core human practice thus becomes critical in evaluating and designing technologies that intervene in the intimate work of memory-making.

\subsection{Point of departure}\label{sec:point_of_departure}
In response to the prevalence of conversational agent systems that aim to automate reminiscence and displace the facilitator, our work instead explores generative personalization as an approach by foregrounding the therapist’s relational, interpretive role in care with support from generative AI tools. We recognize the affective, culturally attuned, and improvisational skills that facilitation requires, particularly under structural constraints. We were motivated to develop a system designed to multiply the capabilities of care workers, rather than replace them.

Given the central role that therapy staff play in facilitating RT, our study sought to actively involve them as experts and users through a participatory design approach. Participatory design is rooted in worker participation in developing computer-based systems for work \cite{kensingblomberg1998}, and has been applied in healthcare to co-design technological interventions for complex multi-stakeholder systems including telemedicine \cite{clemensen2016}, interactive AR-based telepresence activities \cite{ullal2024iterative} and health monitoring systems \cite{cherian2024step}.

In this work, we adopted a RtD process that treats the iterative construction of artifacts as a way of generating knowledge about preferred futures \cite{zimmerman2007}. Within this process, we worked with therapy staff as partners in exploring and defining how AI might reconfigure RT practice through the various prototypes built, rather than treating them as end users of a finished system.
\begin{figure*}[ht]
  \includegraphics[width=\textwidth]{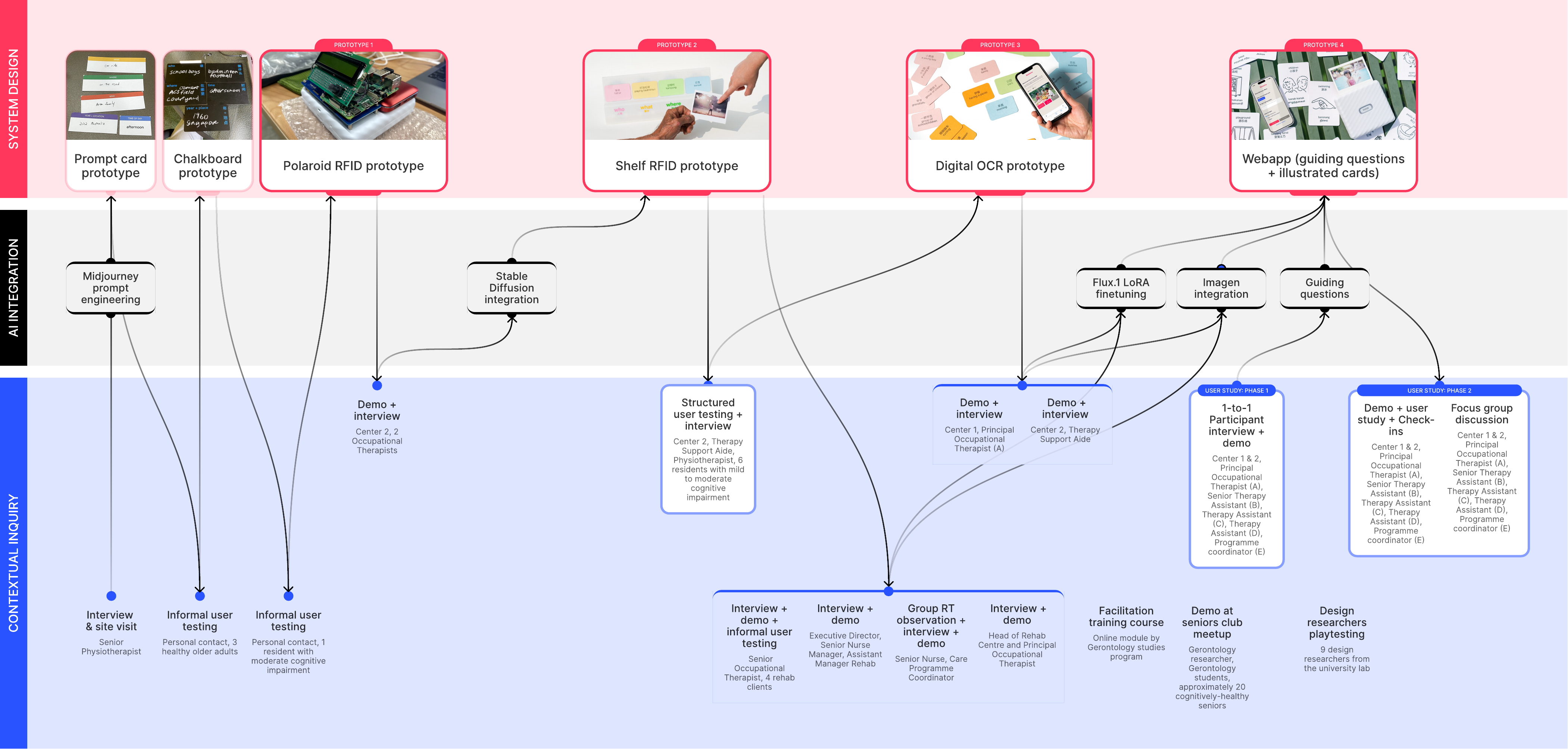}
  \caption{Timeline of RtD process of contextual inquiry, system design and AI integration culminating in a pilot study spanning over two years}
  \Description{A horizontal research-through-design timeline divided into three layers: System Design (top, pink), AI Integration (middle, light gray), and Contextual Inquiry (bottom, blue). Across the top row, 2 small tests and 4 prototypes are shown from left to right with small photos: a prompt-card prototype, a chalkboard prototype, a Polaroid RFID reader (Prototype 1), a shelf-based RFID prototype (Prototype 2), a digital OCR prototype (Prototype 3), and a final webapp with guiding questions and illustrated cards (Prototype 4). Arrows connect these prototypes downward to corresponding user engagements in the Contextual Inquiry layer, such as interviews, informal user testing, structured user testing, group RT observations and multi-session demos at two care centers. The middle AI Integration layer shows when Midjourney prompt engineering, Stable Diffusion integration, Flux.1 LoRA finetuning, and Imagen integration were introduced, with arrows leading to later prototypes and studies. At the bottom, the timeline lists all study touchpoints chronologically, involving physiotherapists, occupational therapists, therapy support aides, rehab clients, program coordinators, and seniors. The flow of arrows illustrates how system design, iterative prototyping, AI model integration and field inquiry influenced one another across two years of development.}
  \label{fig:timeline}
\end{figure*}

\section{Research-through-Design process}\label{sec:research_through_design}
In \anonCountry, RT sits within the holistic long-term care service for people living with dementia and involves many stakeholders and institutions. Rather than defining requirements at the outset, we used the RtD process---specifically Contructive Design Research \cite{koskinen2011}---to generate knowledge through the construction of artifacts in the field. We structured the work around the development, deployment, and study of an AI-assisted tool, \textit{Rememo}, for therapy staff to support their facilitation of RT. Our iterative development of Rememo was informed by contextual inquiry with stakeholders, while also investigating how to coherently integrate generative AI into a tool for RT. These explorations informed the design requirements for each prototype. These prototypes were continually tested with therapy staff culminating in a two-week trial deployment at \anonCareOrganization.

Our work around contextual inquiry, AI integration, and prototype development and study were interwoven throughout the whole RtD process. For presentation clarity and to delineate these separate design research activities, we present our process in the following three sections, Contextual inquiry, System Design, and AI implementation. It is important to note that these categorizations are porous and non-sequential, rather than a rigid three-way partition. \autoref{fig:timeline} illustrates the activities as they occurred chronologically, mapped to three corresponding sections.

Our team comprises a design researcher as the first author and an Occupational Therapist as a co-author, enabling us to combine direct clinical insight with the design processes to develop an \textit{ultimate particular} \cite{thedesignway} for the context. The other authors are HCI design researchers working in the areas of services, healthcare, and tangible interactions, and served as mentors in these respective areas.

Initial field visits were conducted in a sensitizing capacity to better understand the contexts in which RT happens. These included informal engagements at five dementia care facilities in \anonCountry, comprising nursing homes, rehabilitation centers, and senior daycare centers. Due to logistical arrangements and willingness of different partners, we chose to focus on one organization, \anonCareOrganization, for the subsequent user study.

\begin{figure}[b]
  \includegraphics[width=\columnwidth]{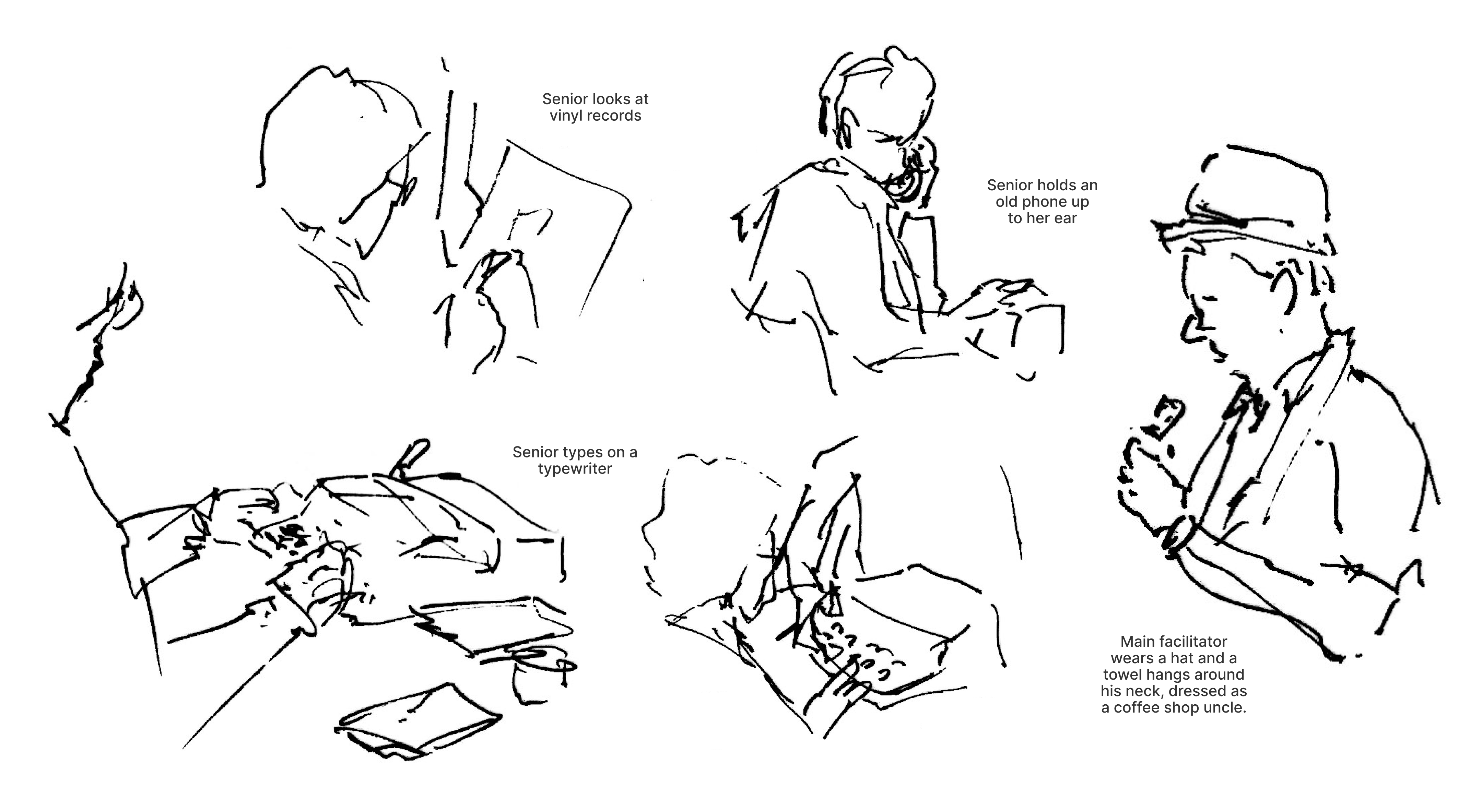}
  \caption{Sketches from field observations of staff and seniors at various care facilities}
  \Description{Collage of simple black line sketches from a reminiscence session. Top left: a senior looks at vinyl records. Top center: a senior holds an old phone to her ear. Bottom left and center: a senior types on a typewriter. Right: the main facilitator, dressed like a coffee shop uncle with a hat and a towel around his neck, stands holding a microphone while addressing participants.}
  \label{fig:sketches}
\end{figure}

During these visits, the first author engaged in ethnographic observations and informal conversations with therapists and therapy teams. Sketches and field notes (\autoref{fig:sketches}) were taken to document the dynamics while maintaining participants’ privacy \cite{kuschnir2016ethnographic}. These observations captured the relational dimension of reminiscence between facilitators and seniors and highlighted the need for tactile cues.

The first author gained exposure to RT by undergoing reminiscence facilitation training by gerontologists and engaging with active aging seniors at a community event. Playtesting was also conducted with designers and HCI researchers from the university lab to get feedback on the usability of prototypes.

These activities laid the groundwork for a subsequent collaboration with \anonCareOrganization\, a private nursing home operator in \anonCountry. In Phase 1 of the research study, we conducted interviews with occupational therapists, program coordinators and therapy aides at two of their long-term residential care facilities to understand facilitation strategies, work challenges, and the infrastructural realities they faced. In Phase 2, a two-week user study of Rememo was conducted with the same participants, detailed in \autoref{sec:user_study}.

\begin{figure*}[h]
  \includegraphics[width=\textwidth]{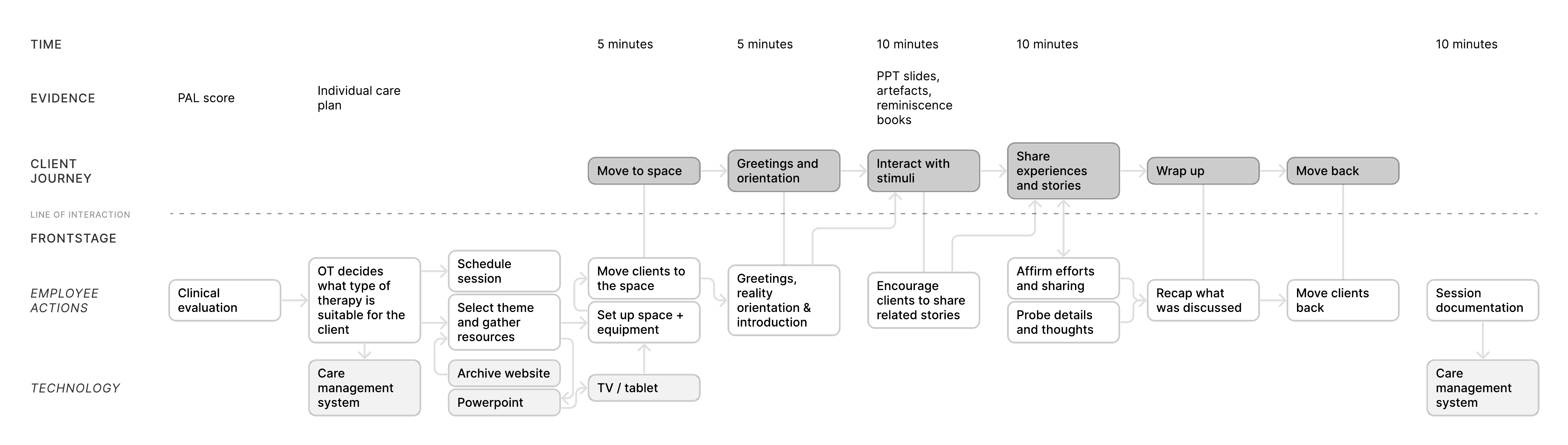}
  \caption{Service blueprint of current RT workflow}
  \Description{Service blueprint of a RT session shown as a horizontal timeline with 5--10 minute stages. Rows from top to bottom: Evidence (PowerPoint slides, artifacts, reminiscence books), Client journey (move to space, greetings and orientation, interact with stimuli, share experiences and stories, wrap up, move back), a line of interaction, Frontstage employee actions (OT decides therapy type, schedule session, select theme and gather resources, set up space and equipment, encourage sharing, affirm efforts and probe details, recap, session documentation), and Supporting technology and processes (care management system, archive website, TV or tablet). Arrows indicate flow across the stages.}
  \label{fig:sb}
\end{figure*}

\subsection{Contextual inquiry}\label{sec:contextual_inquiry}
\subsubsection{Who is the user?}\label{sec:who_is_the_user}
Care technologies are often designed for the care recipient as the primary user (\autoref{sec:related_work}) and we inherited such beliefs too, initially approaching the research focused on care center residents as the primary user. However, through repeated engagements with therapy staff, we came to realize that the success of a reminiscence session often hinges as much on the skill of the facilitator as on the quality of the stimuli used. This was echoed by the reminiscence facilitation training course, which discussed desirable facilitator qualities that hinge on the facilitator’s ability to connect with the client on an emotional and cognitive level.

We also noticed significant variation in the resources available across the 4 care facilities observed. Some were equipped with senior-friendly tablets and curated artifacts, while others relied on printed archives or improvised materials.

Therapy staff also came with varying levels of skill and familiarity with the cultural context. Clients speak different languages in \anonCountry, and non-local care staff who do not speak local languages face significant communication challenges with their clients especially when the work is conversational in nature. 

We therefore scoped our research to focus on supporting therapy staff as the primary user of care systems who negotiate both their relationship with clients and the resources available to them as employees within a care organization. Therapy staff were also involved as co-design partners of the resulting tool, Rememo. It is important to note that in this research, residents were involved as secondary participants through RT sessions facilitated by the therapy staff we worked with. We opted not to work directly with residents in consideration of our focus on care workflows, as well as to respect the professional boundaries and sensitive nature of the work done by our clinical collaborators.

\subsubsection{RT as a fluid practice}\label{sec:rt_as_a_fluid_practice}
Despite being an established intervention, literature has shown that RT lacks standardized protocols \cite{macleod2020}. We found that practices varied widely across care settings, teams, and individual facilitators at \anonCareOrganization. One program coordinator described their sessions as primarily aiming to \quoted{entertain and engage}, while some therapy assistants emphasized achieving the cognitive and behavioral goals set by occupational therapists. These differing motivations determine how a session is conducted. We mapped out the service blueprint of an amalgamated RT workflow which provided a visual map to consider points for design interventions (\autoref{fig:sb}).

The diversity of RT work at \anonCareOrganization\ reinforced our understanding of RT as a fluid and relational practice shaped by the specificities of care, facilitator styles, and client abilities. Therapy staff shared how they adapted sessions, including printing bingo cards bigger for clients with visual impairments, moderating the difficulty of games for each client, and grouping clients by interest and background. This insight guided our design goals toward designing a system that supports therapy staff to respond to different scenarios, rather than prescribing a one-size-fits-all facilitation script.

\subsubsection{From Device to Tools}\label{sec:from_device_to_tools}
\begin{figure*}[h]
  \includegraphics[width=\textwidth]{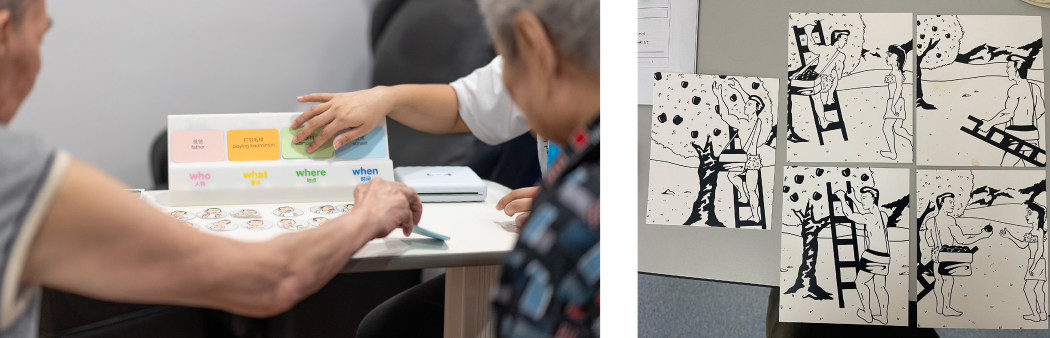}
  \caption{(Left) Therapy support associate places Who, What, Where, When cards on labeled shelves on Prototype 2 during a session with seniors. (Right) Picture cards for story sequencing cognitive test}
  \Description{Two photos. Left: during a user testing reminiscence session, a therapy support associate places “Who, What, Where, When” cards onto labeled shelves on Prototype 2 while an older adult picks up cards on the table. Right: five black and white picture cards arranged for a story sequencing cognitive test, depicting a simple scene with a person on a ladder by a tree across successive frames.}
  \label{fig:wwwwcardsvsstorycards}
\end{figure*}

Viewing RT as a fluid practice with therapy staff at its core led us to reframe our design direction from building an all-in-one device (Prototypes 1 \& 2), to creating a set of tools that can be flexibly embedded into RT practice (Prototypes 3 \& 4) as catalysts for conversation and rich memory work. This shift also changed our approach in crafting prompts for the generative AI engine embedded in our prototypes. Early prototypes (Prototypes 1--3) used a structured Who / What / Where / When format, inspired by narrative structures. However, therapy staff noted that sequencing a coherent story along these dimensions resembled clinical cognitive assessments they used in practice (\autoref{fig:wwwwcardsvsstorycards} right) and could be challenging for residents with more severe cognitive impairments (e.g. in executive functions).

We therefore transitioned to a more flexible thematic system (Prototype 4) comprising broad themes with multiple prompt phrases. This structure enabled open-ended card combinations that accommodate fragmented or non-linear recall. Therapy staff could also enter additional details in free-text to add specific details for each client.

\begin{figure*}[h]
  \centering
  \includegraphics[width=\textwidth]{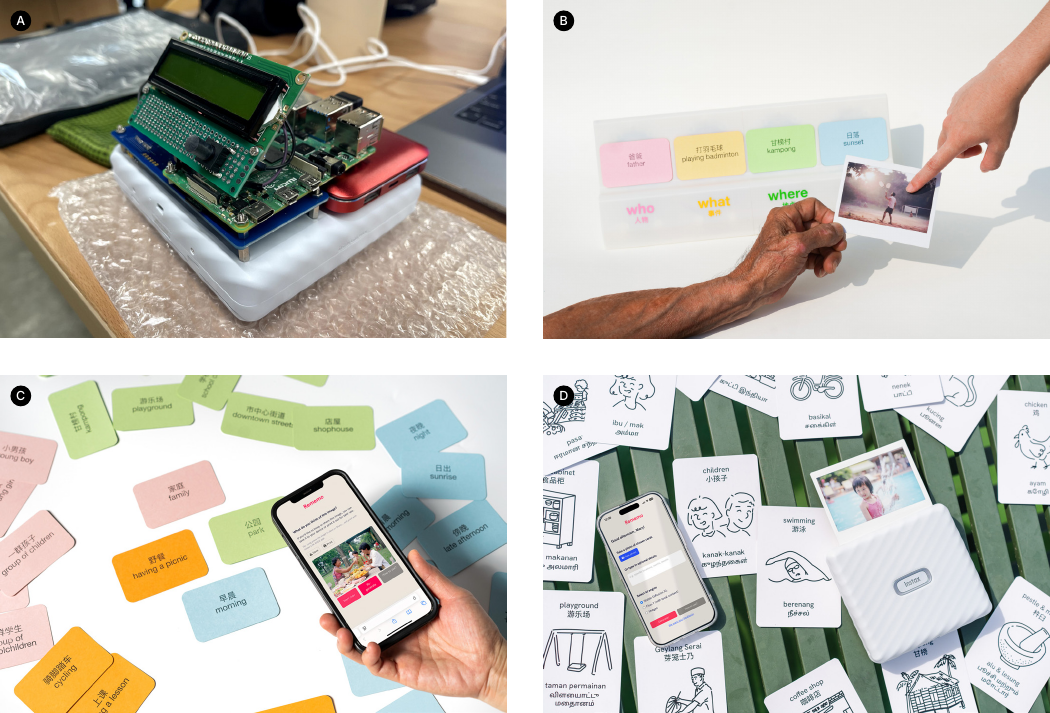}
  \caption{\\A: Prototype 1: Raspberry Pi 4 connected to Adafruit PN532 NFC/RFID reader and LCD display, powered by a battery bank. \\B: Prototype 2: Arduino Uno connected to 4 Adafruit PN532 NFC/RFID readers in a 3D-printed enclosure. \\C: Prototype 3: Mobile webapp prototype. Integrated a camera-based OCR input of physical prompt cards using Google Cloud Vision API. Supports Stable Diffusion XL (SDXL) engine. \\D: Prototype 4: Mobile webapp prototype, with support for 3 image generation engines (SDXL, Flux.1 finetuned with LoRA and Imagen), LLM-generated guiding questions and 128 illustrated prompt cards with multilingual translations and dialect pronunciation support.}
  \Description{Four photos showing successive Rememo prototypes. Photo A, top left: a stack of electronics—a Raspberry Pi board with an LCD and a PN532 NFC/RFID reader mounted on a battery bank. Photo B, top right: a 3D printed “Who, What, Where” shelf with embedded NFC readers; a hand places a small photo while an older adult points. Photo C, bottom left: a smartphone running a mobile web app among colorful prompt cards with bilingual labels. Photo D, bottom right: the mobile web app shown on a phone amid illustrated prompt cards beside a portable photo printer ejecting a print.}
  \label{fig:prototypes}
\end{figure*}

\begin{figure*}[h]
  \includegraphics[width=0.9\textwidth]{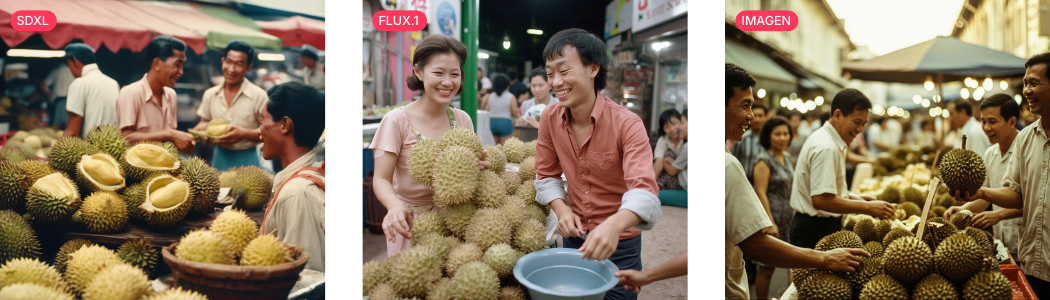}
  \caption{Images generated using Stable Diffusion XL (left), Flux.1 (center) \& Imagen (right) with the prompt \prompted{smiling durian sellers in the market}}
  \Description{Three AI-generated images of smiling durian sellers. Left: painterly look with noticeable distortions; soft facial edges, mushy durian spines, and flat lighting give a stylized rather than photographic feel. Center: desaturated colors and slight noise reminiscent of an early digital camera; geometry is stable with minimal distortions. Right: very clean, high-contrast image with crisp detail and no distortions; anatomy and textures are accurate, but the modern clarity makes the scene feel contemporary rather than archival.}
  \label{fig:generatedimages}
\end{figure*}

\begin{figure*}[h]
  \centering
  \includegraphics[width=0.9\textwidth]{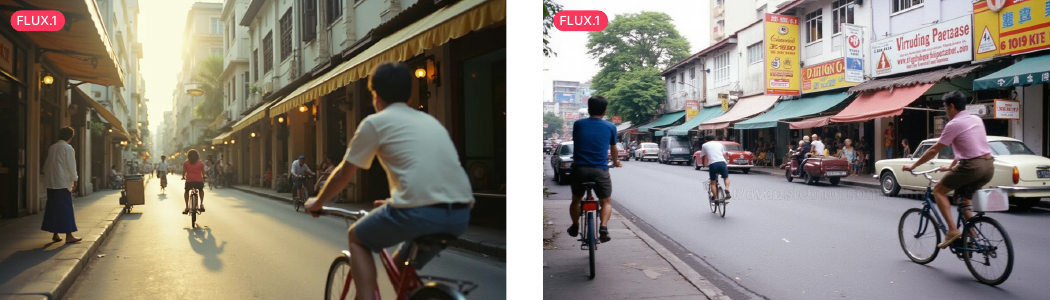}
  \caption{Image generated using Flux.1 dev without LoRA (left) and with LoRA (right) using the same seed and prompt: \prompted{cycling, Chinatown}}
  \Description{Two AI-generated street scenes of people cycling with the prompt “cycling, Chinatown,” contrasted for localization. Left (Flux.1 without LoRA): a sunlit boulevard with European-style facades and wide sidewalks; architecture, signage, and street furniture do not resemble Chinatown and the result reads as a generic city. Right (Flux.1 with LoRA): a row of shophouses with multi-language signboards, five-foot ways, awnings, and street activity that evokes a Southeast Asian street without resorting to cliché props; cyclists and vehicles fit the streetscape, giving a plausible sense of place.}
  \label{fig:loranolora}
\end{figure*}

\subsection{System Design}\label{sec:system_design}
\subsubsection{Tangible vs Digital}\label{sec:tangible_vs_digital}
\autoref{fig:prototypes} summarizes the four iterative prototypes, highlighting how our focus moved from a hardware-based device (Prototypes 1, 2) to a hybrid physical–digital webapp (Prototypes 3, 4) better aligned with practical needs.

Physical interaction with objects for reminiscence support systems have been shown to be favorable for people with dementia \cite{huber2019tangible}, yet they ``must be very robust.’’ 

Our first two prototypes were hardware-based, utilizing microcontrollers connected to NFC/RFID readers. Prototype 1 took inspiration from a Polaroid camera as a memory-capturing device. Users to insert and scan prompt cards, which appends text to a prompt sequentially. This prompt is used to generate an image, which is printed immediately via an instant photo printer. Participants found this interaction unintuitive, and often forgot which cards had already been scanned. In response, Prototype 2 was designed with a more legible physical display. Four card slots were presented side by side which allowed cards to be visibly arranged across the narrative dimensions of Who, What, Where, When. 

Despite the improved affordances of Prototype 2, these custom hardware suffered from fragile electronics and cumbersome set-up requirements. These dependencies meant that the Prototypes 1 \& 2 could only be operated by the research team. These initial breakdowns revealed the dependence of custom hardware on expert maintenance. For therapy staff with existing heavy workloads, operating additional technically demanding equipment was neither realistic nor appropriate to their role \cite{sun2023datawork}. The mismatch between the prototype and labor conditions of care underscored that robustness in this context is not only a matter of material durability, but of aligning with the uneven resources, skills, and time constraints of facilitators.

We responded to these constraints by moving towards a hybrid physical-digital approach that comprise a custom web application that interfaced with physical cards via Optical Character Recognition (OCR). We kept the element of the printed image, maintaining a physical artifact that therapy staff and residents could hold, display, or revisit. 

Through the iterative prototyping process, we reframed tangibility from single robust device to a modular set of tools that therapy staff could flexibly adapt. We reduced hardware complexity through a webapp while retaining tangible media in the form of the printed images. Crucially, the prints and prompt cards were no longer tied to a fixed system but could be used independently or in combination across activities. We observed that therapy staff leveraged this modularity to tailor reminiscence to residents’ needs, demonstrating the shift from monolithic interaction to a versatile medium for dialogue and connection (Section~\ref{sec:from_device_to_tools}).

\begin{figure*}[h]
  \centering
  \includegraphics[width=0.9\textwidth]{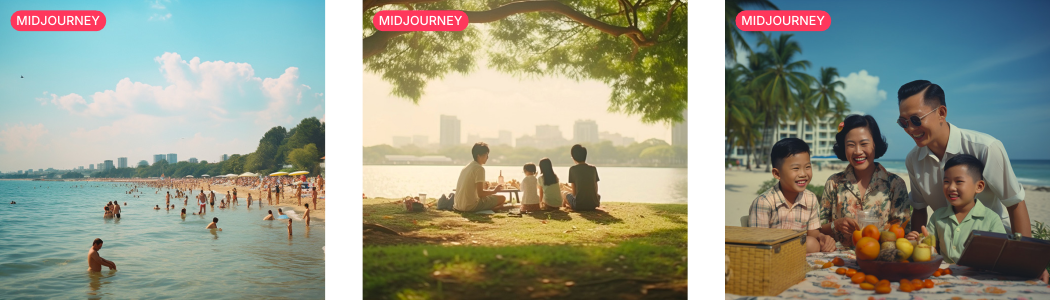}
  \caption{Comparison between Wide context (left), Third-person (center) and First-person (right)}
  \Description{Three AI-generated seaside scenes illustrating composition. Left (Wide context): an expansive establishing shot along the shoreline with many bathers; subjects are small and the environment and horizon dominate. Center (Third-person): a medium wide view from behind a small group seated under a tree, facing the water and distant skyline; the viewer is an external observer and the scene is framed by foliage. Right (First-person): a tight foreground view of a family at a picnic table on the beach; faces and food are close and detailed with minimal background, creating direct engagement with the subjects.}
  \label{fig:widethirdfirst}
\end{figure*}

\begin{figure*}[h]
  \centering
  \includegraphics[width=0.9\textwidth]{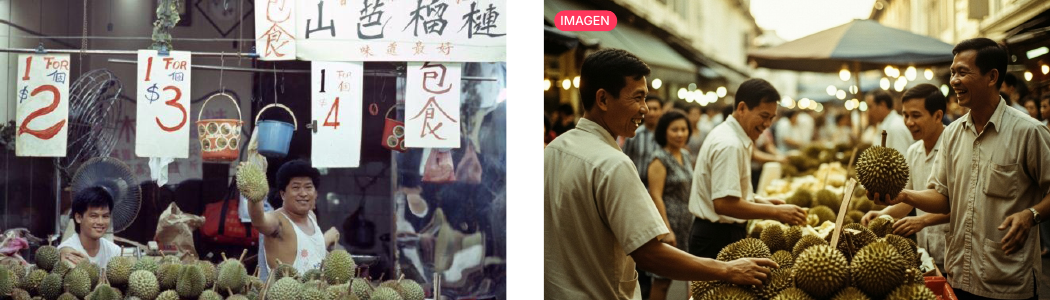}
  \caption{(Left) Durian seller (1985–1999). Photo from G P Reichelt Collection, courtesy of National Archives of \anonCountry\ \cite{nas_durian_sellers}. (Right) Image generated using Imagen with the prompt \prompted{smiling durian sellers in the market}}
  \Description{Comparison of archival and AI-generated durian market scenes, highlighting similarities in subject and differences in photographic texture. Left (archive): frontal view of two vendors in a street stall, with hand-lettered prices on boards, hanging baskets, and piles of durians in crates; the image has muted contrast, slight haze. Right (generated with Imagen): outdoor street market stall with shophouses in the background; several men smile and handle durians across a long counter. The rendering is very clean with high contrast and modern clarity.}
  \label{fig:archivegenerated}
\end{figure*}

\begin{figure*}[h]
  \centering
  \includegraphics[width=0.9\textwidth]{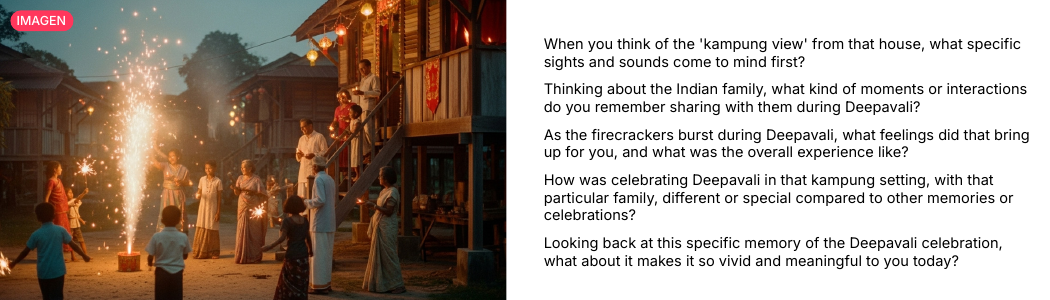}
  \caption{Image generated using Imagen with the prompt card \prompted{Deepavali} and free-text input \prompted{courtres house, kampung view, Indian family, fire crackers} (left) with 5 guiding questions (right)}
  \Description{Left: AI-generated evening scene in a kampung with wooden houses lit by lanterns; an Indian family and neighbors stand outside holding sparklers while a firework bursts in the yard. Adults and children watch together near the house steps. Right: a list of five guiding questions for reminiscence. ``When you think of the 'kampung view' from that house, what specific sights and sounds come to mind first?'', ``Thinking about the Indian family, what kind of moments or interactions do you remember sharing with them during Deepavali?'', ``As the firecrackers burst during Deepavali, what feelings did that bring up for you, and what was the overall experience like?'', ``How was celebrating Deepavali in that kampung setting, with that particular family, different or special compared to other memories or celebrations?'', ``Looking back at this specific memory of the Deepavali celebration, what about it makes it so vivid and meaningful to you today?''}
  \label{fig:generatedimages5qns}
\end{figure*}

\begin{figure*}[h]
  \centering
  \includegraphics[width=\textwidth]{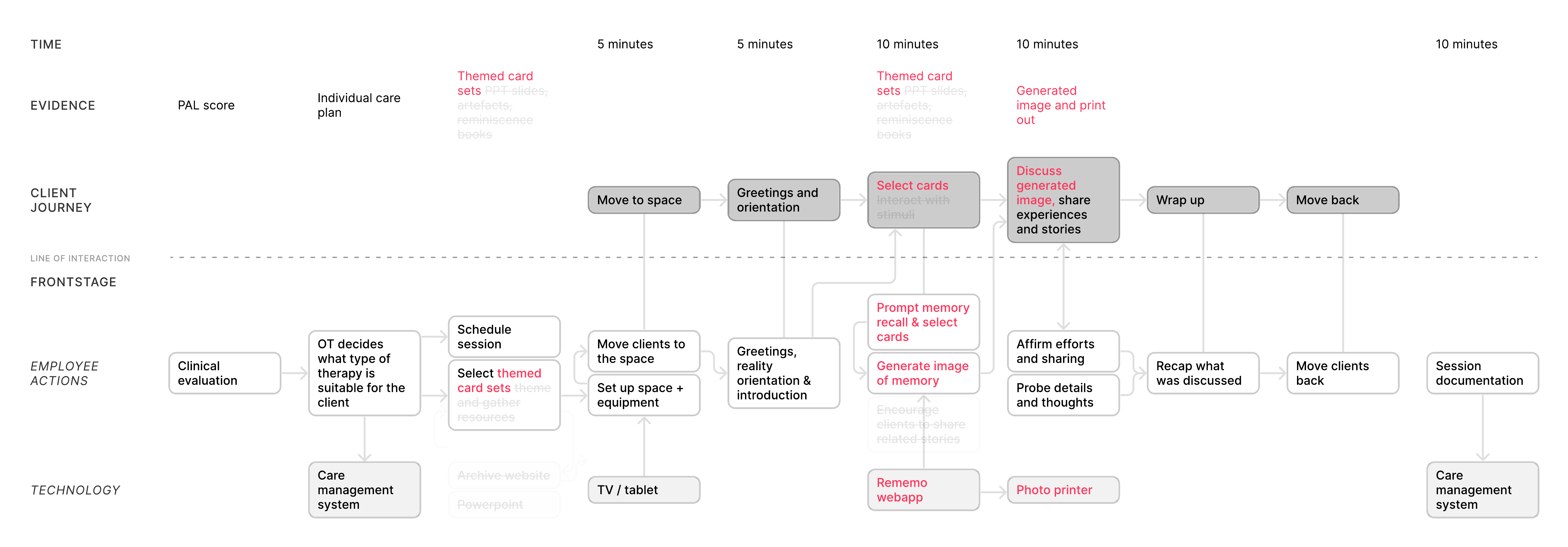}
  \caption{Service blueprint of envisioned RT workflow with Rememo}
  \Description{Service blueprint of the envisioned RT workflow using Rememo. Time runs left to right in 5–10 minute blocks. Rows show Evidence, Client journey, Frontstage line of interaction, Employee actions, and Technology. Magenta items indicate changes introduced by Rememo. Evidence adds themed card sets during preparation and a generated image and printout during the session. Client journey: move to space, greetings and orientation, select cards, discuss the generated image and share experiences, wrap up, move back. Employee actions: schedule session, select themed card sets, move and set up, prompt memory recall and select cards, generate an image of memory with the Rememo webapp, affirm sharing and probe details, recap, document session. Technology supports include care management system, TV or tablet for display, Rememo webapp, and a photo printer. Arrows connect stages to show flow across preparation, interaction, and documentation.}
  \label{fig:envisionedsb}
\end{figure*}

\begin{figure*}[h]
    \includegraphics[width=0.9\textwidth]{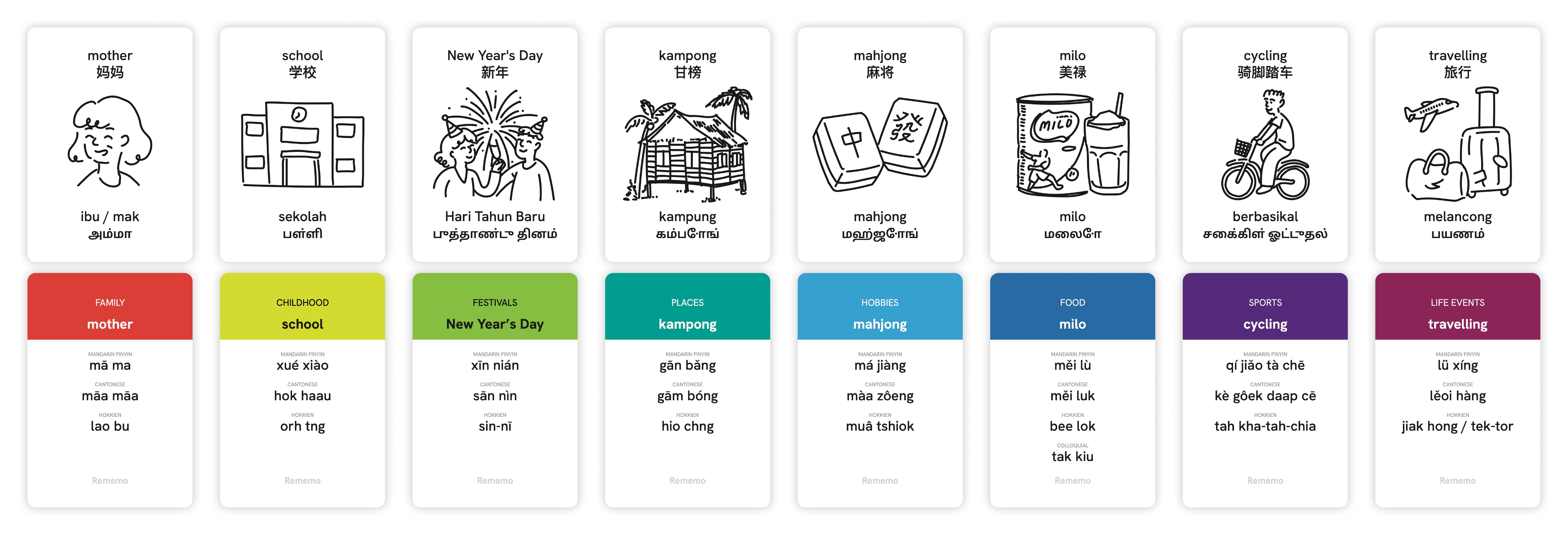}
    \caption{Front and back of a prompt card from each of the 8 categories. Each illustrated prompt card is equipped with translations in English, Mandarin, Malay and Tamil in addition to pronunciation romanization for Mandarin, Cantonese, Hokkien and colloquial terms.}
    \Description{Row of eight illustrated prompt cards from the Rememo set, showing the front and back. The front of each card has a black-and-white line drawing, the term in English, Mandarin, Malay and Tamil. On the back of each card, the term in English is placed within a color band. The color band is coded by category: Family, Childhood, Festivals, Places, Hobbies, Food, Sports, Life Events. Below, there are pronunciation romanization for Mandarin, Cantonese, Hokkien and colloquial terms. Visible cards read: mother, school, New Year’s Day, kampong, mahjong, milo, cycling, and traveling.}
    \label{fig:cards}
\end{figure*}

\subsection{AI Integration}\label{sec:ai_integration}
As discussed in \autoref{sec:related_work}, Generative AI can be used to curate and personalize stimuli for RT, and we used it to elicit personal memories in a way that mirrors conventional reminiscence practices in this research. Specifically, we used image models to generate visual stimuli and large language models to generate guiding questions for RT facilitation.

We continually evaluated and adapted the image generation engines used in our prototypes to achieve optimal visual quality as AI models evolved throughout the course of our work. We tested a range of models, including Midjourney, Stable Diffusion XL, Flux.1, and Imagen, to support different image styles (\autoref{fig:generatedimages}) and provide redundancy in case of platform instability. We eventually excluded Midjourney as it does not support API services which prevents it from being integrated into a custom software application. 

One persistent challenge for using of Generative AI systems in this work is the bias embedded in training data, reflected through Eurocentric aesthetics and culturally generic stereotypes \cite{shav2024}. To address this, we curated a dataset of 317 publicly available archival images depicting past local scenes and used it to fine-tune a Low Rank Adaptation (LoRA) model for the Flux.1 image generation engine. The resulting model produced images with greater cultural specificity and historical texture, while also capturing the visual tone of old photographs familiar to our target users. 

\autoref{fig:loranolora} shows how a LoRA-finetuned model produces a more historically plausible background that aligns with \anonCountry’s architecture, whereas the untuned model yields a more generic scene, even inserting red lanterns to convey a stereotypical Chinatown aesthetic.

To generate engaging visuals that provoked reminiscence, we developed prompts and evaluated outcomes with stakeholders. We initially identified three types of image compositions: Wide context shots that evoke environmental familiarity, Third-person views that allow residents to project themselves or recognize others in the scene, and first-person perspectives to encourage immersion (\autoref{fig:widethirdfirst}). Therapy staff provided feedback that wide context shots would overwhelm the individual with excessive visual detail. Ultimately, we only used third-person and first-person views that made it easier for the individual to place themselves in the scene.

Therapy staff also showed us examples of existing images used in RT, often actual photographs from the past. To match this visual language, we steered prompts and model selection toward photorealistic rendering, explicitly avoiding images that are abstract or highly stylized that could confuse residents (\autoref{fig:archivegenerated}).

Alongside image generation, we also explored generating guiding questions for inexperienced therapy staff to facilitate deeper reminiscence. As discussed in \autoref{sec:labor_realities_of_care_work} and \autoref{sec:therapy_facilitation_as_a_human_practice}, many therapy staff are non-local and lack the cultural understanding that plays into residents’ memories. We paired each generated image with five open-ended guiding questions generated by an LLM (Gemini 2.5 Flash) following a tailored template from the perspective of an experienced reminiscence facilitator (\autoref{fig:generatedimages5qns}). The questions cover topics like the emotions, sensory details, relationships, cultural contexts, and personal significance. The questions are presented as guides that therapy staff can use or take inspiration from, centering human-led facilitation supported by generative AI.

\begin{figure*}[h]
    \includegraphics[width=0.8\textwidth]{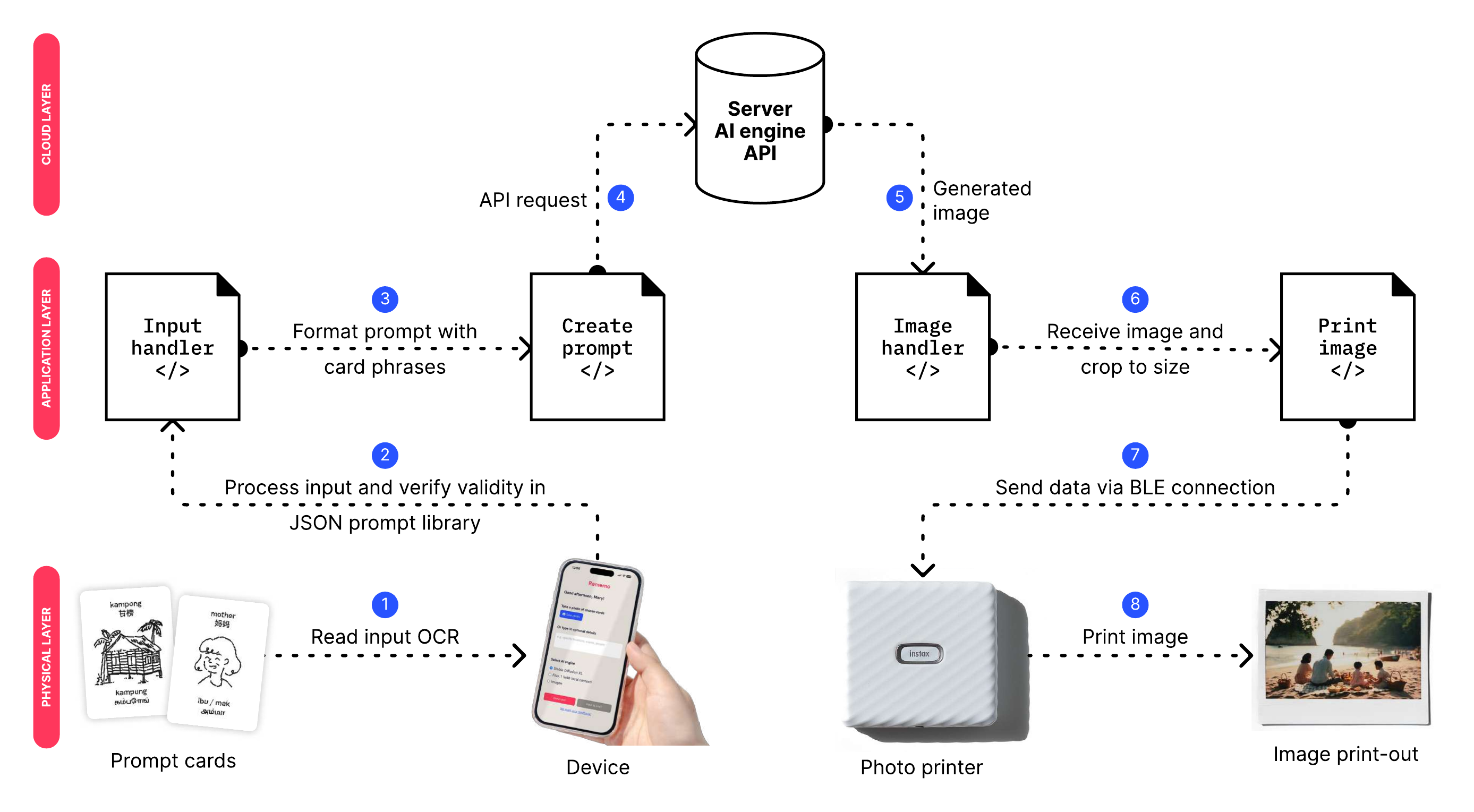}
    \caption{Rememo system architecture diagram}
    \Description{System diagram showing data flow across three layers in 8 steps. Physical layer: prompt cards are scanned with a mobile device and a photo printer produces the final print. Application layer: modules labeled Input handler, Create prompt, Image handler, and Print image process the request. Cloud layer: a Server AI engine API. Numbered steps indicate the flow: (1) the device reads card text via OCR, (2) validates input against a JSON prompt library, (3) formats the prompt with card phrases, (4) sends an API request to the server, (5) generates an image, (6) receives the generated image and crops to size, (7) sends data via BLE to the printer, and (8) prints the image.}
    \label{fig:systemarchitecture}
\end{figure*}

\subsection{Rememo system}\label{sec:rememo_system}
Our iterative RtD process culminated in Rememo, a therapist’s tool that integrates generative AI into RT to support their relational practice through a tangible–digital hybrid interface (Prototype 4). Each component of the system reflects a response to situated constraints---such as resource availability, language barriers, and cognitive variability---while drawing on the therapy staff’s expertise in shaping meaningful client interactions.

\autoref{fig:envisionedsb} shows how we expect Rememo to reshape RT, with differences from current practices highlighted. Before an RT session, the therapy staff selects a relevant theme and brings the corresponding cards, saving time and effort spent gathering materials from reminiscence packages. During the session, clients browse the cards, choosing those that speak to them, and using their selection to generate a personalized image. We envision that this will shift the experience from fixed, preselected material to a visual memory trigger that is co-constructed in the moment.

\begin{figure}[h]
    \includegraphics[width=\columnwidth]{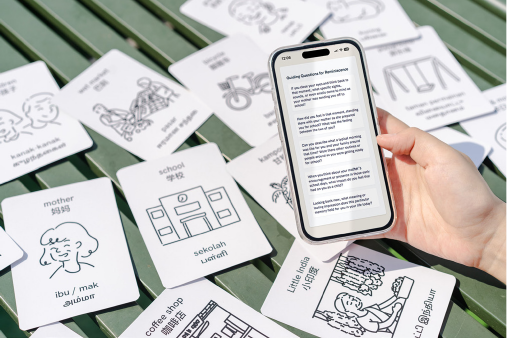}
    \caption{Five LLM-generated guiding questions tailored to selected cards}
    \Description{Hand holding a smartphone that displays Rememo’s guiding questions. The phone rests above a spread of illustrated prompt cards.}
    \label{fig:guidingqns}
\end{figure}

\begin{enumerate}
    \item \textit{\textbf{Themed Prompt Card Sets:}} We designed eight themed sets of illustrated prompt cards (e.g. family, places, life events) to act as physical conversation starters to engage residents. Each card includes translations in \anonCountry’s four official languages, dialects, and common colloquial terms. These cards form the primary input modality. (\autoref{fig:cards})
    
    \item \textit{\textbf{Mobile Web App with OCR Scanning:}} Therapy staff scan selected cards using a mobile device through the web app we developed. We use optical character recognition (OCR) to detect the text on the cards and trigger image generation. This device-agnostic setup enables easy deployment in everyday care environments.

    \item \textit{\textbf{Personalized AI-Generated Imagery:}} 
    The selected card phrases are consolidated and packaged as a prompt (\autoref{fig:systemarchitecture}) for image generation using one of three supported text-to-image engines: SDXL, Imagen, or Flux.1. The generated image is prompted to be culturally familiar to older adults in \anonCountry.
    
    \item \textit{\textbf{Therapist-Moderated Preview and Printing:}} Therapy staff review the generated image and determine its appropriateness. If suitable, the image can be printed using an instant photo printer (Instax Link Wide) which also serves as a tangible keepsake. This step maintains therapist discretion and ensures that images are only shared when evaluated to contribute positively to the session.

    \item \textit{\textbf{Guidance for Novice Therapy Staff:}} Each generated image is accompanied by guiding questions tailored to the card content. These prompts help novice therapy staff ease into facilitation, providing conversational scaffolding without imposing rigid scripts (\autoref{fig:guidingqns}).

\end{enumerate}

\section{User study}\label{sec:user_study}
\subsection{Study design}\label{sec:study_design}
We conducted a deployment-based user study to understand how therapy staff incorporated Rememo (\autoref{fig:prototypes} Prototype 4) into their existing RT sessions, and how it supported or challenged their facilitation process. We were guided by the following research questions (RQs): 

\begin{enumerate}
    \item \textbf{RQ1:} How do therapy staff integrate the Rememo system into their existing reminiscence therapy practices across various client profiles? How is its use similar or different from conventional RT methods?
    \item \textbf{RQ2:} What are the usability and practical considerations when using the Rememo system in therapy sessions?
    \item \textbf{RQ3:} What are the perceived impacts of Rememo as an AI-mediated RT tool? 
    \item \textbf{RQ4:} What kinds of futures do therapy staff envision for AI-enabled RT?
\end{enumerate}

We recruited five care staff from \anonCareOrganization\ (including the second author) as participants in the study. Participants were required to have experience conducting RT sessions in any capacity, with job functions ranging from Occupational Therapist (OT), Therapy Assistant (TA) to Care Program Coordinator. \autoref{tab:participants} details the profiles of the five participants. 

\begin{table*}[h]
  \caption{Demographic Details of Therapy Staff Participants}
  \label{tab:participants}
    \begin{tabular}{ccccc}
      \toprule
      ID & Job designation & Center & Country of Origin & Languages spoken \\
      \midrule
      A & Principal Occupational Therapist & Center 1 & \anonCountry\ (Local) & English, Mandarin \\
      B & Senior Therapy Assistant & Center 2 & \anonCountry\ (Local) & English, Mandarin, Malay \\
      C & Therapy Assistant & Center 2 & \anonCountrytwo & English, Malay, Tamil \\
      D & Therapy Assistant & Center 1 & \anonCountrytwo & English, Tamil\\
      E & Care Program Coordinator & Center 2 & \anonCountrythree & English \\
      \bottomrule
    \end{tabular}
\end{table*}

The study ran across 2 long-term residential care facilities operated by \anonCareOrganization\ in August 2025. The study began with a one-hour prototype demonstration, followed by a two-week period where participants used the tool at their discretion with residents. At the end of each session with Rememo, participants complete a survey form and log session details including their experience with Rememo, resident profiles, and their responses. We conducted a midpoint check-in to address technical and usability issues. At the end of the study, a focus group discussion (FGD) was held to reflect on participants’ experiences, impressions of the tool, and broader questions about the role of AI in memory work. The survey questions and FGD guide can be found in the Supplementary Material. Our study focused primarily on the therapy staff’s experiences and workflows as the main users of Rememo, rather than the care recipient (\autoref{sec:therapy_facilitation_as_a_human_practice}).

\begin{table}[h]
  \caption{Deployment Study Timeline and Activities}
  \label{tab:study-activities}
  \begin{tabularx}{\columnwidth}{lX}
    \toprule
    Phase & Description \\
    \midrule
    Phase 1 & 1-to-1 interviews with participants (1 hour) \\
    Phase 2 & Prototype demonstration with walkthrough of features and card sets (1 hour) \\
             & Self-directed use of Rememo in regular RT sessions \\
             & Midpoint check-in for technical troubleshooting (30 mins) \\
             & Focus group with participants by center (1 hour) \\
    \bottomrule
  \end{tabularx}
\end{table}

A total of 26 sessions with 21 residents were conducted by 5 participants. 19 were one-to-one sessions and 7 were group sessions (held only at Center 1), with group sizes of 2 to 4 residents. The arrangement of group or individual sessions was determined by participants based on the resident’s profile and sociability, as well as center logistics. Sessions lasted 21--55 minutes (M = 32). Solo sessions were typically by the resident’s bedside, though group sessions and some individual sessions occurred in shared spaces such as the Rehab room, dining areas, or corridors.

Residents were recruited based on their Pool Activity Level (PAL) scores\footnote{Pool Activity Level is the standard assessment metric \anonCareOrganization\ used to evaluate the abilities of their residents and tailor the program for their needs. It is a checklist covering 9 elements and residents can be organized into 4 levels: Planned, Exploratory, Sensory and Reflex. \cite{pool1999}}, with inclusion restricted to those assessed as Planned or Exploratory, indicating sufficient ability to follow instructions and express themselves in simple language. 19 residents were assessed as Planned, 1 as Exploratory, and 1 was not recorded.

The resident cohort reflected a broad age range: 4 were 51--60, 6 were 61--70, 7 were 71--80, and 4 were 81--90. Cognitive impairment levels were heterogeneous, with 6 residents showing no impairment, 10 mild impairment, and 5 moderate impairment. These distributions provided a heterogeneous but functionally communicative sample, aligning with the study’s focus on supporting facilitated reminiscence. 

All interviews and FGDs were audio-recorded and transcribed verbatim. Transcripts, sessions surveys, and system usage data were then thematically analyzed with reference to the RQs to surface insights from participants’ use of Rememo.

\subsubsection{Ethics}\label{sec:ethics}
This study was approved by the Institutional Review Board of the first author's institution (\irbone\ \& \irbtwo). All research activities focused on therapy staff as participants who gave their informed consent. No identifiable data were collected from residents. Clinical staff outside of the study obtained informed consent or assent from residents or their next-of-kin on behalf of the research team to minimize unnecessary personal data disclosure and coercion due to the dependent relationship between the participants and residents.
\section{Findings}\label{sec:findings}
\subsection{Sessions conducted}\label{sec:sessions_conducted}
Participants generated 151 images across 26 sessions, of which 50 (33.1\%) were printed onto physical media. Imagen was the most frequently used engine (n = 83, 55.0\%), followed by Flux (n = 53, 35.1\%) and SDXL (n = 15, 9.9\%). Print rates, taken here as a proxy for participants’ assessment of an image’s viability in-session, varied across engines: 34.9\% (29/83) for Imagen, 34.0\% (18/53) for Flux, and 20.0\% (3/15) for SDXL. The full session logs along with participants’ notes are provided in the Supplementary Materials. We summarise how participants incorporated Rememo into their workflow in a modified Service Blueprint in \autoref{fig:modifiedsb}.

\subsection{RQ1 \& 2: Incorporating Rememo into existing RT routines}\label{sec:incorporating_rememo_into_existing_rt_routines}
Overall, participants found Rememo \quoted{easy to use}\footnote{All gray italicized text are participant quotes} including taking a photo of the selected cards and adding free-text input. Minor usability issues such as connecting to the photo printer, saving the generated image, and scanning the cards with OCR were resolved upon clarification.

\subsubsection{Working within the logistical realities of care} 
Rememo eased logistical preparation and making facilitation enjoyable for both therapy staff and residents. Participants described it as a kit where, meaning therapy staff \quoted{don’t really need to think [about] what to talk about.}, addressing a long-standing challenge where less experienced \quoted{TAs (therapy assistants) don’t know what to talk [about].} The guiding questions was valued as a practical aid helping to structure sessions and sustain conversation. Beyond preparation, participants also described the process as enjoyable and enriching, allowing them to \quoted{learn something new, because we are not [from] their [residents] era.} 

\begin{figure}[h]
  \includegraphics[width=\columnwidth]{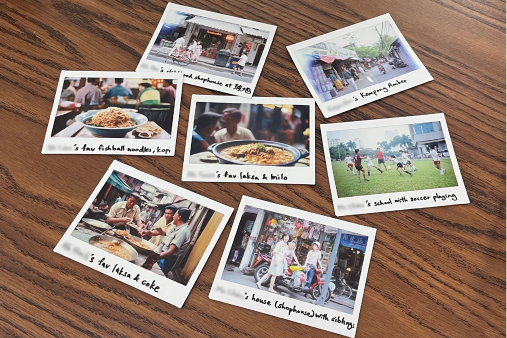}
  \caption{Printed photos from sessions annotated with handwritten captions by participants}
  \Description{Seven small printed Polaroid photos laid out on a wooden table, each with a handwritten caption along the bottom white border. The images include street market scenes, plates of local food, a neighborhood street, and children playing on a field. The captions were written by participants to document resident's memories.}
  \label{fig:printedphotos}
\end{figure}

Physical prints (\autoref{fig:printedphotos}) offered tangibility and gave seniors something to \quoted{touch and feel,} evoking nostalgia and inspiring Participants A and D to compile “life story books” from the prints. 
Yet the cost of printing photos was a concern, as \quoted{normally [the] budget is very low for group therapy [in terms of] resources given by the higher management}. Instead, participants projected generated images onto a TV or tablet, particularly in group sessions as \quoted{in groups, [if] you use the booklet [it’s] very small, [and the therapy staff would] have to go one by one.} This shift highlights the balance between the benefits of tangible media and the practical realities of cost and scalability.

\subsubsection{Rememo as a medium for communication} 
Rememo helped reduce communication barriers, but participants still needed to adapt how they engaged residents. Non-local staff like Participant D \quoted{still have the language barrier} with residents who only speak another language. 
Though the translations on the cards made it \quoted{easy to talk to them}, some residents could not engage with the cards independently. 
Participant C explained, \quoted{some don’t know how to read, [so] we will just read it to them,} often narrowing choices to two or three cards to avoid overwhelming them. Participants also gave feedback that the guiding questions were too complex for residents as \quoted{some [were] too abstract}, and they would \quoted{simplify the question} or \quoted{use our own questions} instead. 

Participants found Rememo more effective for one-to-one sessions than for groups, where diverse personalities and abilities made facilitation more challenging. In \quoted{one to one [sessions, therapy staff] will go deeper} to understand \quoted{what they like, what is their background, what is their interest.} In contrast, group sessions were more challenging to control when residents chose different topics, leading to divergent discussions. To mitigate this, participants pre-selected a common theme on behalf of the group, placing residents with shared experiences or similar cognitive ability together. Residents with \quoted{very short attention span or very low cognitive level} were better suited for individual rather than group sessions. Even common topics such as such as ``childhood’’ and ``school’’ could expose differences, as Participant A observed that residents who had attended university were \quoted{quite proud to share} while \quoted{the rest who don’t have a chance to go to university cannot really talk or not relate.} These findings underscored the tool’s strength in tailoring reminiscence individually, while highlighting that group sessions still required careful curation by the therapy staff of themes that cater to the dynamics within a group of residents.

These limitations and therapy adaptations are not unique to Rememo. Conventional RT literature identifies attention span and the ability to recall and vocalize as prerequisites for reminiscence \cite{dempsey2014reminiscence}. Participation varies when these prerequisites are only partially met. Facilitators currently strive to adapt conventional RT to each client’s abilities so that residents can still take part (\autoref{sec:therapy_facilitation_as_a_human_practice} \& \autoref{sec:rt_as_a_fluid_practice}). Rememo is designed with this same adaptivity in mind through its modular components. We therefore treat these frictions not as fixed limitations of the system or residents, but as areas to be addressed through further testing and iterative refinement.

\begin{figure*}[h]
  \includegraphics[width=\textwidth]{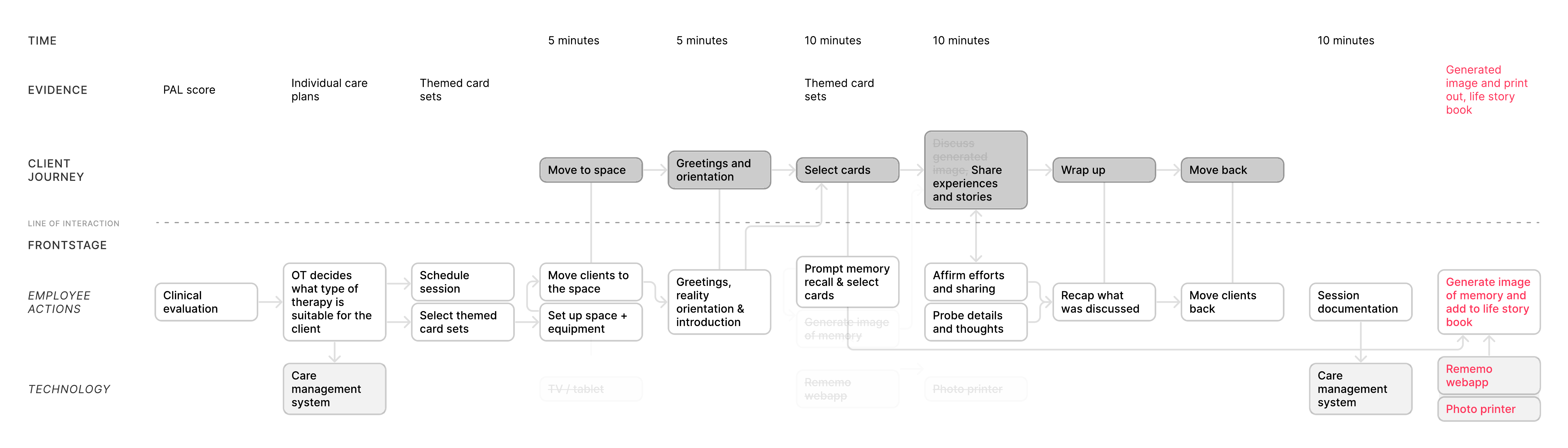}
  \caption{Modified service blueprint shows how Participant A and D adapted to latency by generating images after the session for future use instead of during the session.}
  \Description{Modified service blueprint with time flowing left to right in 5--10 minute blocks. Rows show Evidence, Client journey, Frontstage line of interaction, Employee actions, and Technology. The central session flow remains: move to space, greetings and orientation, select cards, share experiences and stories, wrap up, move back. The change is highlighted on the far right: instead of generating and printing an image during the session, “Generated image and print out” appears after the session. Correspondingly, Employee actions list “Generate image of memory and add into life story book,” and the Technology row shows the Rememo webapp and photo printer in the post-session column. This depicts how Participants A and D adapted to latency by deferring generation to after the session for later use.}
  \label{fig:modifiedsb}
\end{figure*}

\subsection{RQ2 \& 3: Dealing with Generative AI in a therapy flow}\label{sec:dealing_with_generative_ai_in_a_therapy_flow}
\subsubsection{Latency} 
The user study showed that the technical performance of the system directly impacted both usability and the flow of reminiscence sessions. All participants reported long waiting times, \quoted{about half a minute at least}, for image generation. We measured image generation post-deployment and clocked an average of 27 seconds for generations that used card input and 19 seconds for free-text only generations. Latency, which did not emerge as a concern during prototyping nor in demonstrations with participants, proved to be a critical limitation in real-world deployment. Participants described it as \quoted{time-consuming} and disruptive to the conversational rhythm. 

Mismatched or inappropriate generated outputs also required multiple iterations; Participant D, for instance, averaged seven iterations before printing an image. In group settings, conversations often moved on before the photo was generated, with residents \quoted{already [having] changed topic}. During individual sessions, participants managed this latency by \quoted{keep talking} to fill the waiting time. In one session, two of Participant A’s residents grew \quoted{impatient in waiting}, leading to her suggestion to skip image generation during the session and instead use the prompt cards as cues for conversation. 

\subsubsection{Quality of Generated Images:}
Based on our study, we characterize the properties of an AI-generated image for RT that were important to therapy staff:

\begin{enumerate}

    \item \textbf{\textit{Image quality}} refers to the aesthetic fidelity of the output. High-quality images are visually coherent, sharp, and free of distortions. Earlier engines such as SDXL often produced warped figures, while newer models like Imagen largely avoided such distortions.
    \item \textbf{\textit{Prompt adherence}} captures how closely a generated image aligns with the prompt description. 
    Generated images with strong adherence incorporated the all elements within the prompt accurately. Flux often omitted key details or misrendered objects, while Imagen demonstrated stronger adherence when prompts were detailed and descriptive.
    \item \textbf{\textit{Historical accuracy}} was particularly significant in the context of RT. Beyond visual fidelity, participants emphasized alignment with historical contexts, including accurate representations of familiar places, clothing, and objects. Flux was more effective at reproducing the stylistic look of archival photographs, but the AI engines often fell short in consistently capturing the specificity of time and place, which reflected their limited historical grounding in data from \anonCountry.
    
\end{enumerate}

\subsubsection{Dealing with latency \& quality} 
To cope with long loading times, participants at Center 1 began generating images after the session based on the chosen cards and discussions, then saving them for future use rather than on-the-spot printing,  modifying the use of Rememo towards what worked for them (\autoref{fig:modifiedsb}). 

\begin{figure*}[h]
  \includegraphics[width=0.9\textwidth]{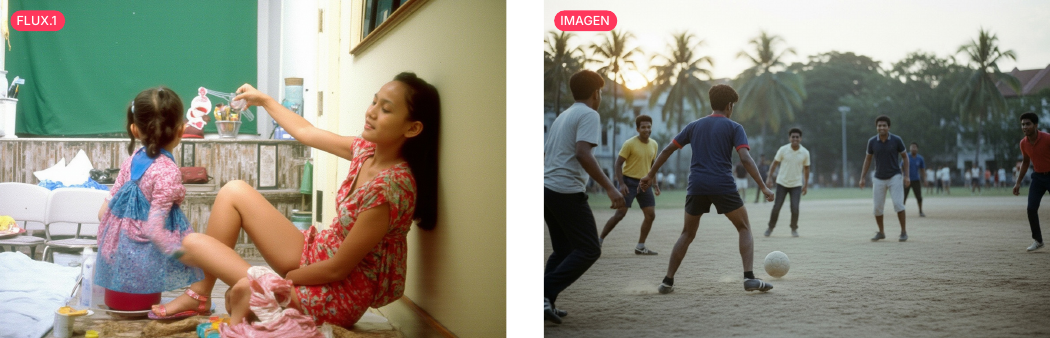}
  \caption{(Left) Image generated using Imagen with the prompt card \prompted{children} and free-text input \prompted{indian 2 daughter...2 grandchildren}, before midpoint check-in. (Right) Image generated using Imagen with the prompt card \prompted{soccer} and free-text input \prompted{played at padang.. with all indian friends..defender..evening time}, after midpoint check-in.}
  \Description{Two AI-generated images contrasted for prompt fidelity and artifacts. Left (pre-check-in, the prompt is ``children'', ``indian 2 daughter… 2 grandchildren''): an adult woman sits on the floor beside a girl wearing a pink and blue dress with unknown distorted objects around. The scene does not match the requested family relations or number of children. Limbs are disjointed and some missing, with other objects blurring into each other, it is unclear what background setting or context of the image. Right (post-check-in, prompt is ``soccer'', ``played at padang.. with all indian friends.. defender.. evening time''): a group of young men play on an open field at dusk. One player adopts a defender stance facing an opponent with the ball. Anatomy, spacing, and interactions are coherent and align with the described roles and setting, yielding higher image fidelity and prompt adherence.}
  \label{fig:prompting}
\end{figure*}

The research team provided Center 2 participants with advice on selecting the appropriate engines for different scenarios and how to write better prompts for image generation during the midpoint check-in. After this, participants reflected that they \quoted{just need to be specific with what we enter on the [system]} and \quoted{to write a story [with] as many [details] as we can, then it’ll really come out.} This was reflected in usage data as the majority of the images produced were through Imagen (54.97\%) over Flux (35.10\%) and SDXL (9.83\%), consistent with Imagen’s stronger prompt adherence and image quality. \autoref{fig:prompting} shows how Participant C evolved her prompting before and after the check-in.

The balance between speed and quality only became visible during deployment, revealing latency as a situated issue of time in the service system. Following Oogjes and Desjardins’ temporal vocabulary \cite{oogjes2024}, the pause that disrupts the flow of reminiscence during image generation can be read as \textit{other-time}, i.e. time when a nonhuman actor---like an AI model---works. While \textit{other-time} was non-issue (and a privilege) for us as system designers testing prototypes in a controlled setting, study participants found it disruptive to the flow of RT. Our check-ins offered prompting guidelines to improve output quality. We also observed participants shifting \textit{when} they used Rememo to account for such \textit{other-time}.
These findings suggest two complementary lenses for working with latency: (1) A \textit{service lens} views other-time as opportunities for therapy staff facilitation, such as chatting or restructuring activities. (2) A \textit{technological lens} seeks faster, more efficient models. We see these lenses as synergistic: workflow supports make waiting productive, while technical improvements shorten or smooth the intervals. Our findings suggest future developments should prioritize supporting the service system while pursuing technological enhancements in parallel.

\subsection{RQ3: How Rememo impacts residents’ recall}\label{sec:how_rememo_impacts_residents_recall}
Participants perceived Rememo as positively affecting residents’ recall and engagement. Many residents were able to recall specific details such as dates, locations, and people, even pointing out differences between the generated image and their memories. Participant C shared that her residents \quoted{really enjoyed this session and they like to keep talking and talking. They're looking forward to the coming sessions… They said this really helps them to remember those days and everything. And they actually love to share with people. So when I go [meet] with them they say, ``Thank you for coming and talking to us and bringing us back to that age''. So they feel it's like something that they want to remember back and they’re happy to share and they want to share more [and] talk more.}

Several residents became more talkative, which indicated benefit to residents’ emotional wellbeing.
Participant D pointed out resident d3, usually \quoted{so quiet} and \quoted{doesn't want to communicate with others} became slightly more engaged in the first session and significantly more in the second session with Rememo. They talked more and even \quoted{listened [to] what other people are thinking} in the second session. Other \quoted{chattier} residents also demonstrated better social interactions with other residents in the session. Participant A noted that a6, previously \quoted{quite chatty} began to \quoted{try to listen to what others are sharing… [such that] he will start to have this interaction among [the residents], so it's no longer one way [sharing, but more] like going to [a] coffee shop.} The continuous conversation demonstrated turn-taking ability by the residents and reduced effort for the facilitator to guide them. 

We surface two possible reasons for this: Resident Agency and Visualization aids.

\subsubsection{Resident agency}\label{sec:resident_agency}
Rememo gave residents a calibrated level of agency to select which memories they wish to discuss through prompt cards. Participant C noted that, in contrast, complete freedom to choose any topic can overwhelm residents, saying \quoted{What you are trying to do? What you want me to do? I don't know.} and \quoted{once they start [to say they] don't know, they will not do.} With the cards, more cognitively able residents could choose from the cards independently while the residents with greater cognitive impairment selected from 2--3 pre-selected cards, sometimes with participants reading the card aloud for residents with visual difficulties. This aligned with existing therapy practice, as Participant E shared about how she encouraged her residents to \quoted{practice choice-making}. The ability to generate new and individually-specific images meant there were \quoted{always new topics to talk about} and participants could \quoted{put everything I want together}. They felt that this individual-relevance puts residents in a positive mood as it was \quoted{something that really related [to] them,} preventing repetition and boredom associated with existing resources.

\subsubsection{Visualization aid}\label{sec:visualization_aid}
Rememo also provided visual aids at every step, from illustrated prompt cards to AI generated images. Participants explained that many clients \quoted{cannot understand verbally [and] rely more on visuals} and Rememo offered an \quoted{actual visual of what they say} that deepened conversations. Participant C noted that though the AI generated images \quoted{cannot be 100\% accurate, but [it’s] almost the same, like almost touching [it]} as her residents reported the images to be about 70--80\% accurate.

Taken together, Rememo supports person-centered care by fostering agency and providing visual aids. It gave residents a sense of control and affirmation, aligning with Kitwood’s emphasis on sustaining personhood \cite{kitwood1997}. At the same time, not all residents benefited equally as those with more severe cognitive impairment or requiring extensive prompting remained at \quoted{status quo}. The broader implication is that interventions like Rememo can go beyond memory recall but also to sustain identity and dignity.

\subsection{RQ4: Futures of AI-enabled RT}\label{sec:futures_of_ai_enabled_rt}
Participants envisioned futures for AI-enabled RT that prioritized speed, historical accuracy, and personalization, so that the technology could keep pace with the dynamic realities of reminiscence. A recurring request was faster image generation to match \quoted{dynamic and spontaneous} conversations, preventing sessions from stalling while waiting for outputs (\autoref{sec:dealing_with_generative_ai_in_a_therapy_flow}). Beyond speed, participants wanted richer and more specific themes. Participant A suggested covering \quoted{their different phases of life from their childhood to teenage to adult... because [at] different phases [of life] they will have different life experiences.} Participant D called for historically grounded prompt cards tied to specific places, figures and national identity that paired with accurate era-specific generated imagery. Center 1 participants described having to \quoted{Google search what [did] the old hospital really look like} and check the AI outputs against actual references. 

Participants also emphasized personalization. Participant A remarked, \quoted{If you can put the real image inside [it] will be better,} suggesting the integration of residents’ own photographs which \quoted{some keep by their bedside} to strengthen recognition and emotional connection. They also imagined expanding beyond static images towards multimedia stimuli with video and sound.

\subsubsection{Negotiating professional boundaries}\label{sec:negotiating_professional_boundaries}
Although participants wanted faster, more accurate, and more personalized AI outputs, they drew clear boundaries around what AI should not do. They rejected the notion of conversational agents as replacements for facilitation, describing \quoted{talking to a robot} as \quoted{sad.} Participant E noted that such systems possess \quoted{no feelings} and lack the emotive personality flair of the facilitator that makes sessions \quoted{interesting.} Participant A similarly reflected, \quoted{For me, if I’m thinking for myself, I wouldn’t really like to talk to a robot,} though she acknowledged that it might still help informal caregivers to \quoted{initiate meaningful conversations.} These reflections highlight an important distinction between AI as a tool to enrich content and communication and AI as a substitute for the relational labor of facilitation.

At the same time, participants envisioned AI-enabled interactive features that could strengthen their practice, particularly for communication barriers. Participant E proposed \quoted{voice control, [whereby] the elderly says something [to] prompt then it will reply,} while Participant C imagined the tool speaking in \quoted{different voices} and \quoted{their dialect} so that even residents who don’t speak the same language could engage more fully. These ideas underscore that therapy staff want AI to extend their reach by easing frictions like speed, accuracy, and language barriers.

\subsection{Limitations}\label{sec:limitations}
While our findings showed promise, we point out some limitations of the study. A key part of the RtD process required the design researchers to collaborate closely with therapy staff over the duration of the study. While this enabled us to generate rich qualitative findings, it also limited the sample size and restricted our research to one national context. Hence, while the findings describe the situated context and perspectives of the participants in nuanced detail, they should not be generalized without caution. Second, the observed signs of improved engagement and recall in residents were encouraging, but the full extent of the therapeutic impacts can only be more thoroughly understood through longitudinal studies and established clinical outcome measures in future studies. Finally, the work is grounded in the context of \anonCareOrganization\ nursing homes in \anonCountry, where infrastructural conditions and cultural practices may differ from other countries. A key challenge in this context is the multilingual, multicultural social fabric, which complicates communication between therapy staff and residents. We addressed this by incorporating language support into the design of Rememo but such features may be less critical in more culturally homogeneous societies. 

\section{Discussion}\label{sec:discussion}
\subsection{AI-in-the-loop}\label{sec:ai_in_the_loop}
Our findings can be categorized into different types of tasks shared between human therapy staff and the Rememo system. We group these tasks across three practical aims: preparing the session, enabling communication, and provoking resonance. For preparation, Rememo enabled new session formats, while therapy staff adapted them to residents’ needs and curated group composition. For communication, Rememo offered language support and guiding questions, while therapy staff steered conversations by engaging with residents’ stories. For provoking resonance, Rememo generated personalized visual imagery that anchored reminiscence, while therapy staff facilitated sharing of personal stories in detail beyond the image. In all, Rememo supplies timely guidance and content, while therapy staff supply judgment, pacing, and sensitivity, so that the overall effect is greater than either could achieve alone.

This division of labor between AI-mediated tools and therapy staff reveal that systems should participate within the therapy workflows rather than act as a standalone agent. This leads us to advocate for an AI-in-the-loop---as opposed to human-in-the-loop---model of therapist-AI collaboration. Human-in-the-loop paradigms are prevalent in AI design, where autonomous systems act while humans intervene only as overseers \cite{settles2009}. In sensitive contexts like dementia care, such delegation risks inappropriate system actions before human correction. For instance, Hsu et al. \cite{hsu2025bittersweet} demonstrates the risks of deploying one-on-one AI technologies for reminiscence with older adults due to the system’s incapacity for emotional sensitivity. Beyond therapy, consumer-facing LLMs have exhibited harmful effects to users’ mental health due to unchecked sycophantic reinforcement \cite{morrin2025delusions}. In contrast, Algorithm-in-the-loop \cite{green2019aitl} and AI-in-the-loop \cite{natarajan2025aitl} frameworks better align with our findings around therapy staff’s needs. 

This orientation is also consistent with the relational nature of reminiscence. Therapy staff adapt to meet people living with dementia at their level rather than follow rigid scripts \cite{macleod2020}. The prototypes we built explored supporting RT in different ways, from structured physical devices to a set of tools for flexible reflection. Therapy staff viewed all as useful depending on the situation. This underscores their skill in shaping the activities to meet client needs. 

AI is capable of generating content, enriching discussions, and bridging language or cultural gaps, but the empathic, improvisational, and affective labor of facilitation remains irreplaceably human. An AI-in-the-loop approach grounds systems in human expertise, with AI supporting at appropriate junctures to augment the human at the heart of memory care \cite{harper2019}. Considering Sun et al.’s work in music therapy, which demonstrated how various musical AI techniques could be matched to different stages of emotion problem-solving \cite{sun2024music}, we envision AI-supported RT as a living library where therapy staff document life histories, source contextually relevant information during conversation, and generate personalized multimedia prompts that enrich reflection and sharing.

\subsection{The case for synthetic memory}\label{sec:the_case_for_synthetic_memory}
The preceding analysis of therapy staff’s desires for speed, accuracy, and personalization, alongside their clear insistence that facilitation remain human-led, foregrounds a critical question: \textit{how should we conceptualize the role of AI-generated imagery in RT?} This was something we struggled with in the beginning of this research as we were concerned with the implications of presenting AI-generated imagery to people living with dementia. Through the study, our views shifted. Rather than positioning synthetic images as deficient imitations of ``authentic’’ photographs within RT, we came to see their value as resonant memory supports that meet a baseline standard for historical plausibility.

Discourse around generative content often emphasizes its artifice and uncertain provenance, devoid of intentionality \cite{reinhart2025} or as dangerous for implanting false memories \cite{pat2025synthetic}. Yet, scholarship on photography has long argued that photographic media cannot be equated with objective truth. Sontag observed that photographs are “not so much an instrument of memory as an invention of it” \cite{sontag1978photography}, that all representational media are interpretive not factual. This distinction is salient in dementia care, where “real” photographs may no longer be recognizable and induce distress rather than support recall. Therapy staff similarly adopt a strengths-based approach in their practice, avoiding correcting people living with dementia but instead gently encouraging them to share their perceptions without prejudice.

From this perspective, the key issue is how therapy staff balance resonance and accuracy, often leaning toward resonance once a basic level of plausibility has been met. Research in cognitive science indicates that human memory is reconstructive rather than archival, more akin to a telephone game than a fixed record \cite{bridgePaller2012, bridgeVoss2014}. Astell et al.’s work showed that the use of generic photos to elicit reminiscence sparked a greater range and depth of personal stories than personal photos which functioned as a “memory test for labels” \cite{astell2010}. In our study, therapy staff demonstrated that images need not be 100\% accurate to stimulate conversation, recognition, and engagement. Instead, even at \quoted{70--80\% accuracy} as assessed by Participant C’s residents (\autoref{sec:visualization_aid}), they were still able to resonate with the content and share memories. The therapeutic goal is therefore not the retrieval of veridical memory, but affective connection, narrative flow, and opportunities for social interaction. Therapy staff thus checked historical plausibility when referring to specific places or periods. But for more general memories, the historical accuracy receded in importance as compared to its resonance with clients’ experiences.

We therefore propose reframing synthetic imagery as an asset rather than a liability for reminiscence. By treating AI-generated images as handles on memory and a resource to facilitate co-construction of stories, Rememo foregrounds the pragmatic and therapeutic function over their documentary status. In doing so, we extend existing discourse on technology-mediated reminiscence by arguing that the synthetic nature of generative media can be deliberately leveraged for person-centered care by accelerating the provision of stimuli, while leaving interpretation and meaning-making firmly in human hands.

\section{Conclusion}\label{sec:conclusion}
In this work we presented Rememo, a therapist-centered tool that integrates generative AI into RT developed through a two-year RtD process. We showed how contextual inquiry and design evolved in response to technical constraints, human infrastructure and cultural specificities. Our deployment study demonstrated how our tool supported therapy staff by easing preparation, bridging communication barriers, and provoking recall and sharing by seniors, while revealing technical and organizational frictions such as latency and cost constraints. 
From these findings, we discuss their implications on the future of designing AI-assisted RT. Our proposition of Rememo as a tool sparked new imaginaries by therapy staff on how AI might enable their work, whether by constructing extensive libraries of personal histories or more engaging multimedia stimuli. These aspirations were underscored by the crucial positioning of AI as a partner in care work, extending the paradigm of AI-in-the-loop to foreground relational labor over human replacement. By centering therapy staff as expert users, Rememo demonstrates how AI can be appropriated to empower rather than replace, sustaining the human relationships at the heart of memory. From these insights, we call for future human-centered AI research for care to move beyond automation, and toward designs that integrate the social, cultural, and affective dimensions.
\begin{acks}
We thank all participants and residents involved for their time and contributions. We are grateful to the rehabilitation teams at \anonCareOrganization\ for their extended support throughout the project. We also thank the reviewers for their thoughtful feedback. This research was funded by NUS Start Up Grant A-0010125-00-00. 
\end{acks}

\bibliographystyle{ACM-Reference-Format}
\bibliography{bib}

\end{document}